\documentclass[fleqn,usenatbib]{mnras}

\usepackage{newtxtext,newtxmath}

\usepackage[T1]{fontenc}
\usepackage{ae,aecompl}

\usepackage{graphicx}	
\usepackage{amsmath}	

\usepackage{amssymb}	
\usepackage{bm}
\usepackage{booktabs}
\usepackage{cleveref}
\usepackage{natbib}

\newcommand{\secpoint}{\mbox{$''\mskip-7.6mu.\,$}}
\newcommand{\angstrom}{\mbox{\normalfont\AA}}

\newcommand{\fesc}{$f_{\rm esc}$}

\newcommand{\jwst}{{\it JWST}}
\newcommand{\hst}{{\it HST}}
\newcommand{\muv}{$\rm M_{\rm UV}$}

\title[UV slopes of early galaxies with JADES]{{The UV Continuum Slopes of Early Star-Forming Galaxies in JADES}}

\author[M. W. Topping et al.]{Michael W. Topping$^{1}$\thanks{michaeltopping@arizona.edu},
Daniel P.\ Stark $^{1}$, 
Ryan Endsley $^{2}$, 
Lily Whitler $^{1}$, 
\newauthor
Kevin Hainline  $^{1}$,
Benjamin D.\ Johnson $^{3}$, 
Brant Robertson $^{4}$, 
Sandro Tacchella $^{5,6}$, 
\newauthor
Zuyi Chen $^{1}$, 
Stacey Alberts $^{1}$, 
William M.\ Baker $^{5,6}$, 
Andrew J.\ Bunker $^{7}$, 
\newauthor
Stefano Carniani $^{8}$, 
Stephane Charlot $^{9}$,
Jacopo Chevallard $^{7}$,
Emma Curtis-Lake  $^{10}$, 
\newauthor
Christa DeCoursey $^{1}$, 
Eiichi Egami  $^{1}$, 
Daniel J.\ Eisenstein $^{3}$, 
Zhiyuan Ji $^{1}$, 
\newauthor
Roberto Maiolino $^{5,6,11}$, 
Christina C.\ Williams $^{12}$, 
Christopher N.\ A.\ Willmer $^{1}$, 
\newauthor
Chris Willott $^{13}$, 
Joris Witstok  $^{5,6}$ \\
$^{1}$ Steward Observatory, University of Arizona, 933 N. Cherry Avenue, Tucson, AZ 85721, USA\\
$^{2}$ Department of Astronomy, University of Texas, Austin, TX 78712, USA\\
$^{3}$ Center for Astrophysics $|$ Harvard \& Smithsonian, 60 Garden St., Cambridge MA 02138 USA\\
$^{4}$ Department of Astronomy and Astrophysics, University of California, Santa Cruz, 1156 High Street, Santa Cruz, CA 95064, USA\\
$^{5}$ Kavli Institute for Cosmology, University of Cambridge, Madingley Road, Cambridge, CB3 0HA, UK\\
$^{6}$ Cavendish Laboratory, University of Cambridge, 19 JJ Thomson Avenue, Cambridge, CB3 0HE, UK\\
$^{7}$ Department of Physics, University of Oxford, Denys Wilkinson Building, Keble Road, Oxford OX1 3RH, UK\\
$^{8}$ Scuola Normale Superiore, Piazza dei Cavalieri 7, I-56126 Pisa, Italy\\
$^{9}$ Sorbonne Universit\'e, CNRS, UMR 7095, Institut d'Astrophysique de Paris, 98 bis bd Arago, 75014 Paris, France\\
$^{10}$ Centre for Astrophysics Research, Department of Physics, Astronomy and Mathematics, University of Hertfordshire, Hatfield AL10 9AB, UK\\
$^{11}$ Department of Physics and Astronomy, University College London, Gower Street, London WC1E 6BT, UK\\
$^{12}$ NSF’s National Optical-Infrared Astronomy Research Laboratory, 950 North Cherry Avenue, Tucson, AZ 85719, USA\\
$^{13}$ NRC Herzberg, 5071 West Saanich Rd, Victoria, BC V9E 2E7, Canada
}

\begin{document}

\label{firstpage}
\pagerange{\pageref{firstpage}--16}
\maketitle

\begin{abstract}
The power-law slope of the rest-UV continuum ($f_{\lambda}\propto\lambda^{\beta}$) is a key metric of early star forming galaxies, providing one of our only windows into the stellar populations and physical conditions of $z\gtrsim10$ galaxies. 
Expanding upon previous studies with limited sample sizes, we leverage deep imaging from JADES to investigate the UV slopes of $179$ $z\gtrsim9$ galaxies with apparent magnitudes of $m_{\rm F200W}\simeq26-31$, which display a median UV slope of $\beta=-2.4$.
We compare to a statistical sample of $z\simeq 5-9$ galaxies, finding a shift toward bluer rest-UV colors at all \muv{}.
The most UV-luminous $z\gtrsim9$ galaxies are significantly bluer than their lower-redshift counterparts, representing a dearth of moderately-red galaxies within the first $500~$Myr.
At yet earlier times, the $z\gtrsim 11$ galaxy population exhibits very blue UV slopes, implying very low impact from dust attenuation.
We identify a robust sample of $44$ galaxies with $\beta\lesssim -2.8$, which have SEDs requiring models of density-bounded HII regions and median ionizing photon escape fractions of $0.51$ to reproduce.
Their rest-optical colors imply that this sample has weaker emission lines (median $m_{\rm F356W}-m_{\rm F444W}=0.19$ mag) than typical galaxies (median $m_{\rm F356W}-m_{\rm F444W}=0.39$ mag), consistent with the inferred escape fractions.
This sample consists of relatively low stellar masses (median $\log(\rm M/M_{\odot})=7.5\pm 0.2$), and specific star-formation rates (sSFR; median $=79~\rm Gyr^{-1}$) nearly twice that of our full galaxy sample (median sSFR$=44~\rm Gyr^{-1}$), suggesting these objects are more common among systems experiencing a recent upturn in star formation. 
We demonstrate that the shutoff of star formation provides an alternative solution for modelling of extremely blue UV colors, making distinct predictions for the rest-optical emission of these galaxies.
Future spectroscopy will be required to distinguish between these physical pictures.
\end{abstract}

%
%
\section{Introduction}
\label{sec:intro}

The slope of the rest-UV continuum (parameterized as $\beta$ where f$_\lambda \propto \lambda^\beta$) provides a valuable diagnostic of star forming galaxies. 
At the highest redshifts accessible with current telescopes (i.e., $z\gtrsim 11$), the UV continuum is often the only detectable part of the spectrum, with the slope providing a rare clue as to the nature of the earliest systems.  
The  intrinsic UV spectral slope is set by the properties of the massive star populations (e.g., metallicity, age) with additional contributions from the spectral shape of nebular continuum emission at very young ages. 
The presence of dust then reddens the intrinsic spectrum to its observed form. For the majority of star forming galaxies, dust plays the dominant role in driving variations in the observed UV colors \citep[e.g.,][]{Wilkins2011}, with stellar properties \citep{Calabro2021} and ionized gas conditions \citep[e.g.,][]{Chisholm2022} having an important but second-order effect.

Deep imaging surveys with the {\it Hubble Space 
Telescope (HST)} have allowed UV continuum slopes to be
computed for large samples of star forming galaxies in
the reionization era \citep[e.g.,][]{McLure2011, Finkelstein2012, Dunlop2012, Rogers2013, Bouwens2014,Bhatawdekar2021}. 
These studies have indicated that $z\simeq 7-8$ galaxies are considerably bluer than those at $z\simeq 2-3$ \citep[e.g.,][]{Dunlop2012,Bouwens2014}, with reionization-era UV slopes often having $\beta \lesssim -2$. 
Dust is likely the primary factor driving this observed redshift evolution in UV slopes, with $z\simeq 7-8$ galaxies facing less attenuation than those at later epochs. 
Some investigations have found a trend toward bluer UV slopes at lower UV luminosities, with the faintest sources (M$_{\rm{UV}}\simeq -18.0$) at $z\simeq 7-8$ exhibiting UV slopes ($\beta \simeq -2.4$) that approach the intrinsic values expected for 
stellar populations and nebular continuum emission \citep[e.g.,][]{Cullen2017}. 
These results potentially indicate a trend between dust attenuation and UV luminosity which has significant implications for the integrated star formation rate density and may hint at a relationship between luminosity and metals in reionization-era systems. 
However, others have argued that the UV slope distribution in the reionization era is entirely consistent with a population average of $\beta \simeq -2$, with 
photometric scatter driving the range of UV slopes 
seen in the galaxy population \citep[e.g.,][]{Cullen2022}.

At yet earlier epochs ($z\gtrsim 9$) we may expect to see additional evolution in the distribution of UV slopes. 
If the massive stellar populations evolve significantly toward lower metallicities at higher redshift, we may find a shift blueward in colors. On the other hand, if the light-weighted stellar population ages become very young, we may expect 
nebular continuum emission to slightly redden UV slopes. And finally, if dust has yet to build up in even the most luminous galaxies at $z\gtrsim 9$, there may be no observed trend between UV luminosity and UV color, with all galaxies showing mostly un-attenuated spectral energy distributions (SEDs).
Unfortunately {\it HST} has never been able to address these questions on its own.  
Not only are its photometric samples too small at $z\gtrsim 9$  \citep[e.g.,][]{Bhatawdekar2021}, but its cutoff at wavelengths redder than the H-band precluded photometric measurements in the wavelength range (2-3$\mu$m) needed for establishing UV colors at these redshifts. 
The combination of {\it HST} and {\it Spitzer} photometry 
provided a first window on UV slopes at $z\simeq 10$, \citep{Wilkins2016, Tacchella2021}, but such efforts were limited to very small samples given the sensitivity of {\it Spitzer} imaging.

The launch of the {\it James Webb Space Telescope} (\jwst{}; \citealt{Gardner2023}) has quickly revolutionized the study of galaxies at $z\gtrsim 9$. 
Thanks to the unprecedented sensitivity of the Near Infrared Camera (NIRCam; \citealt{Rieke2005, Rieke2023nircam}) at 2-5$\mu$m, it is now possible to investigate the distribution of UV slopes in large samples of galaxies at $z\gtrsim 9$.
The first {\it JWST} imaging campaigns led to a number of investigations into the UV slopes of $z\gtrsim 9$ galaxies \citep{Topping2022, Furtak2022, Nanayakkara2022, Bouwens2022, Cullen2022}. 
A key emerging result from these early programs is the striking commonality of blue UV slopes, with typical values of $\beta \leq -2.0$, now extended to the earliest times \citep[e.g.,][]{Robertson2023,Curtislake2023,ArrabalHaro2023}.
The UV continua of these objects provide powerful insight to the physical conditions of galaxies at the highest redshifts, such as low impact from dust and young stellar population ages.  
This advancement has further allowed the identification and confirmation of extremely blue galaxies \citep[e.g.,][]{Topping2022}, whose existence was difficult to establish with \hst{} alone.

The origin of galaxies with $\beta \lesssim -3$ has tested galaxy evolution models since their emergence among $z\gtrsim7$ samples of faint objects \citep{Bouwens2010, Labbe2010, Ono2010}.
These extreme systems have long proven challenging to explain with standard model treatment of emission from stars and gas \citep[e.g.,][]{Robertson2010, Wilkins2011, Zackrisson2013}.
While many physical explanations have been provided, one promising solution has been the result of significant leakage of Lyman Continuum (LyC) radiation \citep[e.g.,][]{Raiter2010b, Zackrisson2017, Plat2019}.
Several observational campaigns have explored this phenomenon in the local Universe, where direct confirmation of ionizing photon leakage is possible  \citep[e.g.,][]{Yamanaka2020, Flury2022b, Chisholm2022}.
Recently, \citet{Kim2023} measured UV slopes on small scales within a confirmed LyC leaker at $z=2.37$ and demonstrated that indeed the UV continua of the leaking regions are extremely blue and consistent with little-to-no nebular continuum. 
Early \jwst{} programs have enabled the identification and study of this class of objects into the distant Universe \citep[e.g.,][]{Topping2022, Furtak2022}; however progress has been stifled due to low sample statistics and uncertainties due to photometric scatter.

In this work, we utilize imaging from the \jwst{} Advanced Deep Extragalactic Survey  \citep[JADES;][]{Eisenstein2023} to address several key shortcomings of current UV slope analyses.
Imaging from JADES has already been demonstrated as an effective tool for identifying and discerning the properties of high-redshift galaxies \citep[e.g.,][]{Hainline2023,Robertson2023,Tacchella2023,Endsley2023}.
We leverage the JADES data to explore the UV slopes for a statistical sample of galaxies at $z\sim5-14$ and covering a large dynamic range of UV luminosity, enabling us to explore trends among the high-redshift galaxy population.
Using this large parent sample, we identify a significant number of extremely blue ($\beta\simeq-3$) galaxies with robust UV slope measurements, which otherwise may suffer from cosmic variance among smaller samples \citep[e.g.,][]{Topping2022}.
We place the observed SEDs of this blue sample in context of models that allow for the escape of LyC photons, and those with complex star-formation histories (SFH).
Finally, we discuss the galaxy properties implied by these two physical pictures, and the different observed quantities that may be able to distinguish them.

This paper is organized as follows.
Section~\ref{sec:data} describes the data and reduction techniques, as well as the sample selection and methods to derive galaxy property.
In Section~\ref{sec:results} we present the UV slopes of our sample and discuss trends with UV luminosity and redshift evolution.
Section~\ref{sec:discussion} presents the selection of a sample of objects with extremely blue UV slopes, followed by a discussion of their properties and exploration of alternative interpretations of the data.
Finally, we conclude with a summary of our key conclusions in Section~\ref{sec:summary}.
Throughout this paper we use AB magnitudes \citep{Oke1984}, and assume a cosmology with $\Omega_m = 0.3$, $\Omega_{\Lambda}=0.7$, $H_0=70 \textrm{km s}^{-1}\ \textrm{Mpc}^{-1}$. 
We additionally adopt solar abundances from \citet[][i.e., $Z_{\odot}=0.014$]{Asplund2009}.

\begin{figure*}
    \centering
    \includegraphics[width=1.0\linewidth]{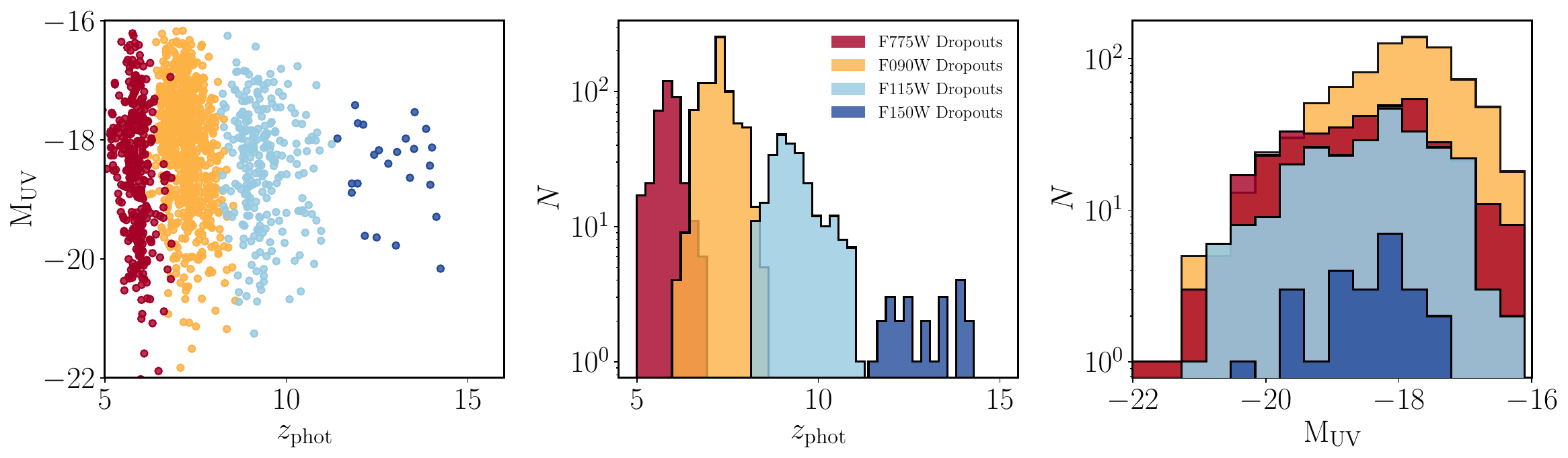}
    \caption{Galaxy demographics for the different samples used in this analysis, and described in Section~\ref{sec:sample}. In each panel we display the ACS/F775W dropout sample (red), NIRCam/F090W dropout sample (orange), NIRCam/F115W dropout sample (light blue), and NIRCam/F150W dropout sample (dark blue). The left panel illustrates the UV luminosity vs. photometric redshift for the four samples derived from the SED fitting that is described in Section~\ref{sec:properties}. Individual histograms of $z_{\rm phot}$ and \muv{} comparing the four samples are presented in the center and right panels, respectively.}
    \label{fig:sample}
\end{figure*}

%
%
\section{Data and Sample}
\label{sec:data}

\subsection{Imaging and Photometry}
We utilized imaging data obtained as part of the JADES collaborative NIRCam+NIRSpec GTO program.
While an in depth description of the full dataset and survey design are presented in \citet{Eisenstein2023}, we provide a brief description here.
The primary imaging dataset was obtained using \jwst{}/NIRCam within the Great Observatories Origins Deep Survey field \citep{Giavalisco2004} South (GOODS-S: RA=$53.126$ deg, DEC=$-27.802$ deg) and North (GOODS-N: RA=$189.229$ deg, DEC=$+62.238$ deg)
obtained during the first year of \jwst{} science operations.
The full observational footprint was observed in four short-wavelength (SW) filters (F090W, F115W, F150W, F200W), and four long-wavelength (LW) bands (F277W, F356W, F410M, F444W).
Data was also obtained in the F335M medium-band filter across a portion of the field, comprising $\sim81\%$ of the full observational footprint.

The JADES NIRCam imaging used in this work were obtained between October 2022 and February 2023, and reductions of these data have been described in several recent publications \citep{Rieke2023, Endsley2023, Robertson2023, Tacchella2023,Williams2023}. 
For this analysis, we employed final mosaics and photometric catalogs produced following the steps discussed in \citet{Endsley2023}, for which we provide a brief description here.
Briefly, the raw uncalibrated exposures frames were first processed through the JWST calibration pipeline utilizing updated reference files based on in-flight performance. 
The resulting images were then corrected for unwanted artifacts, such as wisps and `snowballs' \citep[e.g.,][]{Rigby2023, Bagley2023b}. 
Following this, each image was flat-fielded using custom sky flats and flux calibrated, after which a correction for $1/f$ noise is implemented.
The calibrated images were combined into final mosaics on a $0\secpoint03$ pixel scale, and aligned to the Gaia-EDR3 reference frame.
Finally, all of the mosaics were convolved to the PSF of the F444W filter following the procedures outlined in \citet{Endsley2022-ceers}.
The JWST/NIRCam dataset was supplemented by public archival imaging available in the Hubble Space Telescope Legacy Fields \citep{Illingworth2013, Whitaker2019}.  
Specifically, we utilized mosaics in the GOODS-S field obtained in the HST/ACS F435W, F606W, F775W, F814W, and F850LP filters.
As with the JWST/NIRCam imaging mosaics, the archival HST images were shifted into a consistent astrometric reference frame based on the Gaia-EDR3 astrometry, and placed on the same pixel scale as the \jwst{}/NIRCam mosaics.

The source catalog was derived by first constructing a detection image composed of an inverse variance weighted stack of the F200W, F277W, F335M, F356W, F410M, and F444W NIRCam mosaics.
Sources were subsequently identified from this detection image using \texttt{Source Extractor} \citep{Bertin1996}.
Photometric fluxes for each source were computed in Kron \citep{Kron1980} apertures in each filter, with uncertainties derived by placing copies of a large number of apertures among blank regions of the mosaic surrounding the object, and taking the standard deviation of the distribution of measured fluxes.
These observations obtained as part of the main JADES program reach a typical point source depth ($5\sigma$) of $\sim28.5$ mag in the broadband filters across the entire survey footprint, increasing to a depth of $\sim30$ mag in the deepest region of the mosaics.
Further details regarding the extraction of photometry can be found in \citet{Endsley2023}.

\subsection{Sample Selection}
\label{sec:sample}
The selection of high-redshift galaxies in JADES has yielded an abundant sample of galaxies at $z>8$ identified based on their photometric redshifts \citep{Hainline2023}.
In this work, we used a set of dropout selection criteria to identify galaxies from $z\sim5$ to $z\sim 14$.
Below, we compare the samples constructed from these two methodologies.
Our selections consist of F775W, F090W, F115W, and F150W dropouts, which yield samples of galaxies roughly falling within redshift windows of $z\sim5-7$, $z\sim6.5-8.5$, $z\sim8.5-11$, and $z\sim11-14$.
A detailed description of these selections is provided in \citet{Endsley2023} and Whitler et al. (in prep.), however we provide a brief description here.

Starting with the selection of the highest redshift galaxies, an initial selection of galaxies was obtained for the F150W dropout sample using the following criteria:
\begin{equation}
\begin{cases}
\rm S/N_{\rm F090W, F115W} < 2 \\
m_{\rm F150W} - m_{\rm F200W} > 1.3\\
m_{\rm F200W} - m_{\rm F356W} < 1.0\\
m_{\rm F150W} - m_{\rm F200W} > m_{\rm F200W}-m_{\rm F356W}+1.3
\end{cases}
\end{equation}

At slightly lower redshift, we selected galaxies within $z\sim8-11$ as those that satisfy the F115W dropout criteria:
\begin{equation}
\begin{cases}
\rm S/N_{\rm F090W} < 2 \\
m_{\rm F115W} - m_{\rm F150W} > 1.3\\
m_{\rm F150W} - m_{\rm F277W} < 1.0\\
m_{\rm F115W} - m_{\rm F150W} > m_{\rm F150W}-m_{\rm F277W}+1.3
\end{cases}
\end{equation}

The lowest redshift selection that uses a \jwst{}/NIRCam filter as a dropout band are the F090W dropouts, which were selected as following:
\begin{equation}
\begin{cases}
m_{\rm F090W} - m_{\rm F115W} > 1.3\\
m_{\rm F115W} - m_{\rm F200W} < 1.0\\
m_{\rm F090W} - m_{\rm F115W} > m_{\rm F115W}-m_{\rm F200W}+1.3
\end{cases}
\end{equation}

In addition, each of these selections required non-detections ($\rm S/N<2$) in ACS/F435W, F606W, and F814W, and optical $\chi^2$ \citep{Bouwens2015} of $\chi^2_{\rm opt} < 5$ calculated using the three ACS bands. Finally, a $\rm S/N>3$ detection is required in at least two \jwst/LW bands, with a constraint of a $\rm S/N>5$ detection in F200W.  

Finally, we selected a sample composed of F775W dropouts, intended to obtain galaxies at a redshift of $z\sim5-7$. This sample was constructed using the following criteria:
\begin{equation}
\begin{cases}
m_{\rm F775W} - m_{\rm F090W} > 1.2\\
m_{\rm F090W} - m_{\rm F150W} < 1.0\\
m_{\rm F775W} - m_{\rm F090W} > m_{\rm F090W}-m_{\rm F150W}+1.2
\end{cases}
\end{equation}

Following this initial selection, we employed several steps to clean each of these samples. 
First, we visually inspected the photometry and images of each source in the samples. 
From this visual inspection, we removed objects that were clear artifacts.
This included objects identified as diffraction spikes, hot pixels, objects coincident with the detector edge, and residuals left over from the cosmic ray removal.
We further removed objects from our sample that were significantly contaminated by flux from a bright neighbor, such that the flux from the candidate could not be separated out.
We also fit BayEsian Analysis of GaLaxy sEds \citep[BEAGLE][]{Chevallard2016} models to each of the candidate galaxies using the model setup described in the following section.
A further cut was applied based in the inferred photometric redshifts, such that we require a significant fraction of the total probability ($>80\%)$ lying at high redshift.
This high redshift cutoff was set at $z=4$ for the F775W dropouts, and $z=6$ for the F090W, F115W, and F150W dropouts.
In total, this cleaning process resulted in a sample of $364$, $656$, $232$, and $24$ objects satisfying the F775W, F090W, F115W, and F150W dropout criteria and subsequent photometric redshift cutoff. 
We note that our dropout samples have significant overlap with the sample of $z>8$ galaxies presented in \citet{Hainline2023}, which selected high-redshift galaxies from JADES based on their photometric redshifts as estimated from EAZY \citep{Brammer2008}.
We find that $92\%$ of our sample at $z>8$ (corresponding to the selection criteria of \citealt{Hainline2023}) are in common with their selection.
The remaining $8\%$ of the sample comprise objects at the edge of the dropout selection windows where the completeness is low, or very faint objects that are near the JADES detection limit.

\begin{figure*}
    \centering
    \includegraphics[width=1.0\linewidth]{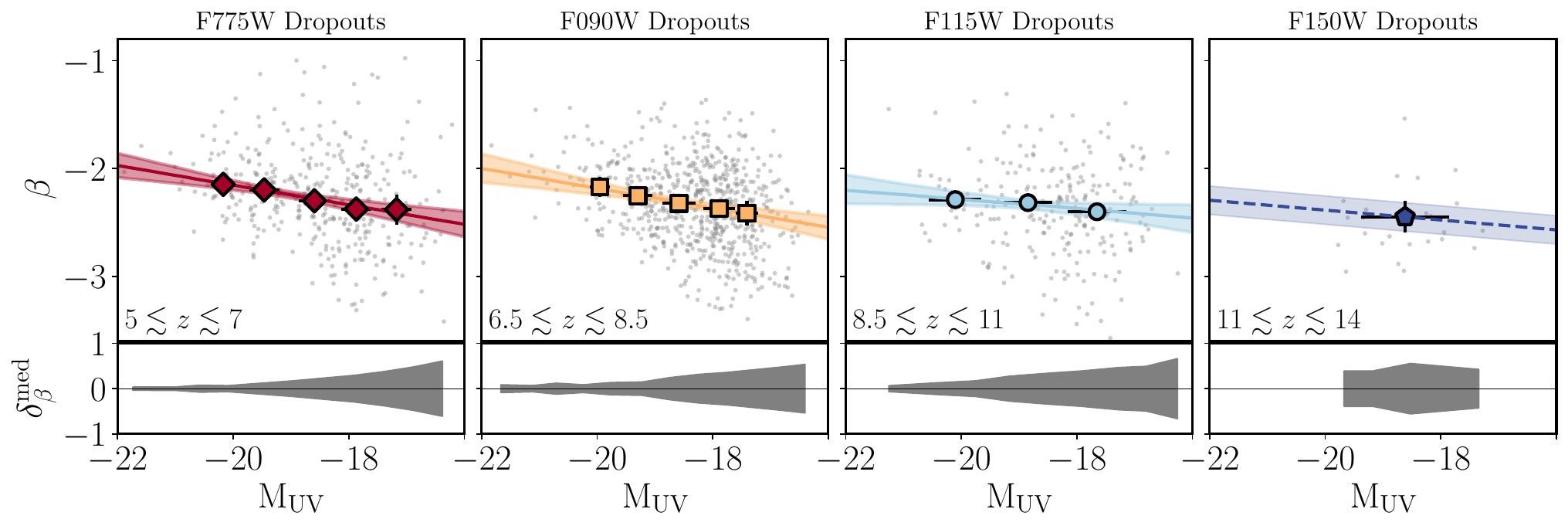}
    \caption{UV slopes of each object falling within our selections. Demographics of each dropout selection sample are provided in Table~\ref{tab:betas}, and values for individual bins are given in Table~\ref{tab:betabins}. Linear fit parameters and uncertainties are presented in Table~\ref{tab:fits}. The slope of the linear fit applied to the F150W dropout sample is fixed to that derived for the F115W dropouts. To distinguish this sample, we display the fit as a dashed line. The lower panel of each plot displays the median uncertainty in the UV slopes of individual objects as a function of UV luminosity.}
    \label{fig:allslopes}
\end{figure*}

\subsection{Measured and Inferred Properties}
\label{sec:properties}
Physical properties for our two high-redshift samples were obtained using a comprehensive suite of SED models.
For all such properties inferred from SED modelling, we adopted values and uncertainties from the median and inner 68th percentile of the posterior probability distribution, respectively.
We derive a fiducial set of model fits obtained using BEAGLE.  These models employ stellar templates from the updated models of \citet{Bruzual2003} that have been post-processed to self-consistently include the effects of nebular emission following the procedures outlined in \citet{Gutkin2016}.
We employed a constant star-formation history (CSFH) with a galaxy age that was allowed to vary from 1 Myr to the age of the Universe with a uniform prior in log space.
We assumed that the stellar and gas-phase metallicities are the same, and adopt uniform priors in log space for metallicity and ionization parameter defined over the ranges of  $\log(Z/Z_{\odot})\in[-2.2, 0.3]$ and  $\log(U)\in[-4,-1]$, respectively.
These models implement the effects of dust using the SMC dust law \citep{Pei1992} as it has been demonstrated to accurately describe the observed properties of high-redshift galaxies \citep[e.g.,][]{Bouwens2016, Reddy2018}.  
Within the models, the effects of dust are parameterized by a $\tau_{\rm V}$, which is allowed to vary from 0.001 to 5 with a log-uniform prior.
We allowed redshift to vary with a uniform prior within the range $z\in[0,20]$.
Finally, we assumed a fixed value for the dust-to-metal mass ratio of $\xi_d=0.3$.

We display the demographics of our samples in Figure \ref{fig:sample}.
This figure presents the UV luminosities (\muv{}) and photometric redshifts inferred from the best-fit BEAGLE models.
The average demographics for each of our dropout samples are presented in Table~\ref{tab:betas}.
We find median UV luminosities corresponding to \muv{}$=-18.41$, $-18.16$, $-18.24$, and $-18.61$.
By virtue of their selection, we find median photometric redshifts of $z_{\rm phot}=5.86$, $7.28$, $9.41$, and $12.02$ for the F775W, F090W, F115W, and F150W dropout samples, respectively.
These same samples span redshift ranges of $5.05-6.84$, $6.07-8.59$, $8.20-11.25$, and $11.40-14.25$.
It is clear that there is some overlap in the redshifts spanned by different dropout samples, however we ensured that each individual galaxy was not identified in more than one sample.

We measured UV slopes for each object in the sample by fitting a power law ($f_{\lambda} \propto \lambda^{\beta}$; \citealt{Calzetti1994}) to fluxes spanning the rest-UV portion of the SED.
The set of filters that cover rest-UV fluxes depend on the redshift of the object, and are chosen to avoid contamination from Ly$\alpha$ emission.
For objects in our sample at $z<5.4$, we measured the UV slope using the F090W, F115W, and F150W fluxes. 
For galaxies with photometric redshifts at $5.4\leq z < 7.4$, we utilized the F115W, F150W, and F200W filters.
At yet higher redshifts of $7.4\leq z <9.8$ and $9.8\leq z<13.2$, we measured UV slopes from the F150W, F200W, F277W, and F200W, F277W, F356W filters, respectively.
Finally, for the few galaxies in our sample at $z\geq13.2$, F277W, F356W, and F444W were used.
For each galaxy we perturbed the photometry and re-measured the UV slope many times, and set the $1\sigma$ uncertainty according to the inner 68th percentile of the $\beta$ measurements.

\begin{table}
\begin{center}
\renewcommand{\arraystretch}{1.2}
\begin{tabular}{cccccc}
\toprule
Selection &  $N$ & $z^{\rm med}_{\rm phot}$ & $\rm M_{\rm UV}^{\rm med}$ & $\beta_{\rm med}$  \\
\midrule
F775W Dropouts &  $364$ & $5.86$ & $-18.41_{-1.29}^{+0.96}$   & $-2.26^{+0.03}_{-0.03}$  \\
F090W Dropouts &  $656$ & $7.28$ & $-18.16_{-1.04}^{+0.76}$   & $-2.32^{+0.03}_{-0.02}$ \\
F115W Dropouts &  $232$ & $9.41$ & $-18.24_{-1.35}^{+0.85}$   & $-2.35^{+0.04}_{-0.04}$  \\
F150W Dropouts &  $24$ & $12.02$ & $-18.61_{-0.91}^{+0.81}$   & $-2.48^{+0.12}_{-0.14}$  \\
\bottomrule
\end{tabular}
\end{center}
\caption{Demographics of the four dropout samples, including median photometric redshifts, \muv{}, and UV slope.}
\label{tab:betas}
\end{table}

\begin{table}
\begin{center}
\renewcommand{\arraystretch}{1.2}
\begin{tabular}{cccccc}
\toprule
Selection &  $N$ & $z^{\rm med}_{\rm phot}$ & $\rm M_{\rm UV}^{\rm med}$ & $\beta_{\rm med}$  \\
\midrule
F775W Dropouts &  $46$ & $6.01$ & $-20.24_{ -0.49 }^{+ 0.23 }$   & $-2.12^{+0.06 }_{-0.06 }$ \\
&$71$ & $5.93$ & $-19.47_{ -0.24 }^{+ 0.26 }$   & $-2.20^{+0.05 }_{-0.04 }$\\
&$89$ & $5.82$ & $-18.59_{ -0.26 }^{+ 0.26 }$   & $-2.30^{+0.05 }_{-0.05 }$\\
&$98$ & $5.72$ & $-17.87_{ -0.24 }^{+ 0.23 }$   & $-2.38^{+0.07 }_{-0.07 }$\\
&$43$ & $5.85$ & $-17.19_{ -0.23 }^{+ 0.26 }$   & $-2.38^{+0.12 }_{-0.13 }$\\
\midrule
F090W Dropouts &$53$ & $7.57$ & $-19.95_{ -0.16 }^{+ 0.14 }$   & $-2.16^{+0.06 }_{-0.08 }$\\
&$86$ & $7.36$ & $-19.27_{ -0.26 }^{+ 0.21 }$   & $-2.25^{+0.04 }_{-0.04 }$\\
&$160$ & $7.32$ & $-18.57_{ -0.29 }^{+ 0.26 }$   & $-2.33^{+0.04 }_{-0.04 }$\\
&$221$ & $7.27$ & $-17.89_{ -0.27 }^{+ 0.25 }$   & $-2.37^{+0.04 }_{-0.04 }$\\
&$136$ & $7.20$ & $-17.40_{ -0.16 }^{+ 0.15 }$   & $-2.41^{+0.10 }_{-0.10 }$\\
\midrule
F115W Dropouts & $37$ & $9.31$ & $-19.99_{ -0.55 }^{+ 0.37 }$   & $-2.28^{+0.06 }_{-0.07 }$\\
&$78$ & $9.26$ & $-18.93_{ -0.39 }^{+ 0.31 }$   & $-2.33^{+0.09 }_{-0.08 }$\\
&$117$ & $9.31$ & $-17.98_{ -0.33 }^{+ 0.42 }$   & $-2.41^{+0.07 }_{-0.07 }$\\

\midrule
F150W Dropouts &  $24$ & $12.02$ & $-18.61_{-0.91}^{+0.81}$   & $-2.48^{+0.12}_{-0.14}$  \\

\bottomrule
\end{tabular}
\end{center}
\caption{Median UV slopes for each dropout sample divided into bins of UV luminosity. Uncertainties on \muv{} and $\beta$ are derived using the Monte-Carlo bootstrap method described in Section~\ref{sec:properties}.}
\label{tab:betabins}
\end{table}

%
%
\section{UV slopes of reionization-era galaxies}
\label{sec:results}

In this section we leverage the imaging depth and area from the JADES program to explore the UV slopes of a statistical sample of $z>9$ galaxies, allowing us to place meaningful constraints across a wide range of galaxy properties.
We first discuss the UV slope distributions of each of the four dropouts samples individually, and explore trends between UV slope and UV luminosity at each redshift in Section~\ref{sec:UVslopes}.
In Section~\ref{sec:zevol} we combine the UV slope measurements of each of the dropout samples and constrain the redshift evolution over $5\lesssim z \lesssim 14$.

\begin{figure*}
    \centering
    \includegraphics[width=1.0\linewidth]{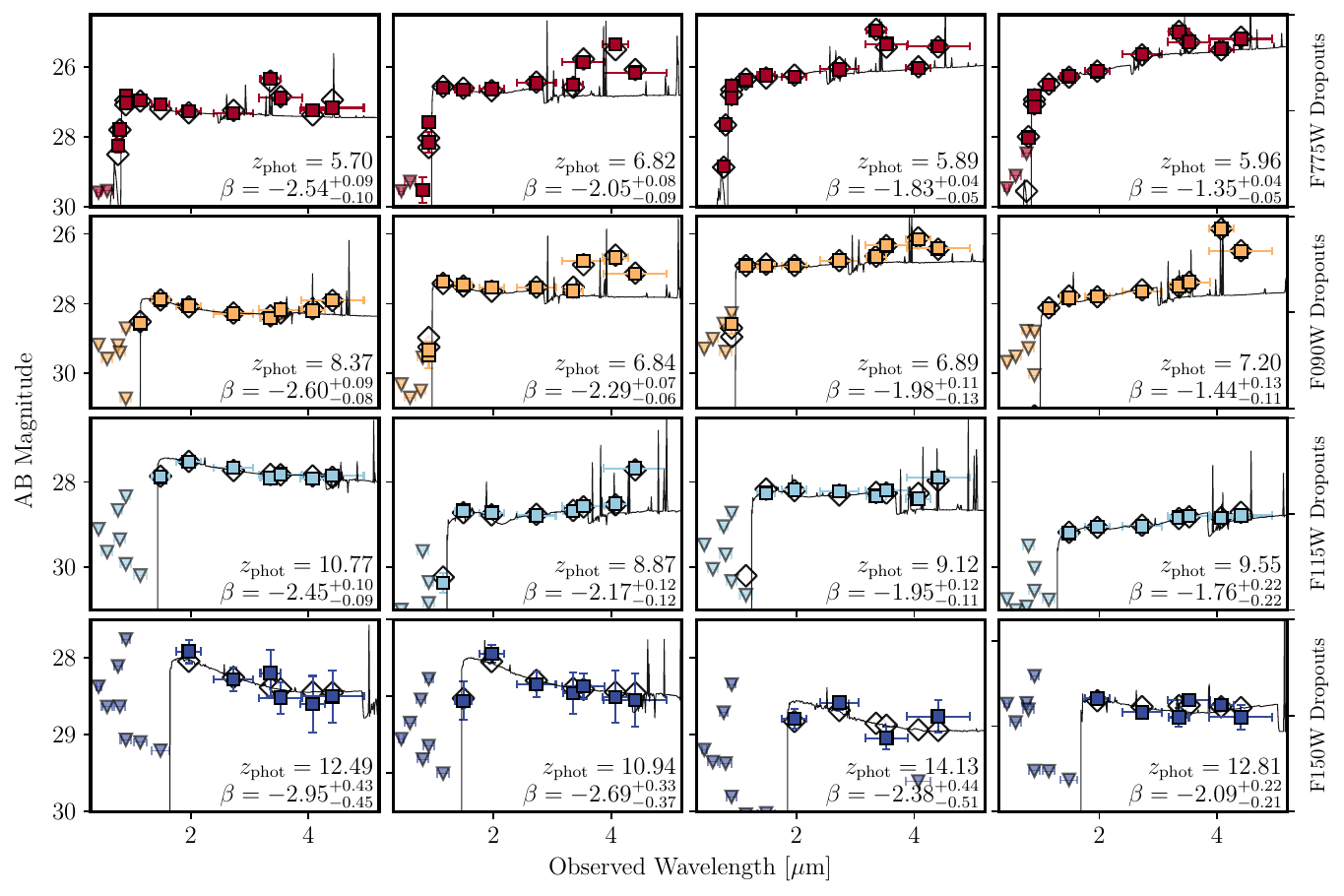}
    \caption{Example objects from each dropout selection, ordered from the lowest redshift sample (F775W dropouts; top row) to the highest redshift sample (F150W dropouts; bottom row), which are color-coded based on their dropout selection as in Figure~\ref{fig:sample}. Empty diamonds represent the photometric fluxes predicted by the best-fit model, and colored squares and triangles are measurements and upper limits, respectively.  We display objects spanning a range of UV slopes within each selection, with bluer galaxies in the left column and redder galaxies in the right column. We list the measured UV slopes with uncertainties in addition to best-fit photometric redshifts for each object.}
    \label{fig:sedexamples}
\end{figure*}

\subsection{UV slope distribution and \muv{} trends}
\label{sec:UVslopes}
We present the UV slopes of our sample comprising $1276$ galaxies identified at $z\sim5-14$ across four dropout selections in Figure~\ref{fig:allslopes}.
The median UV slopes for each dropout selection are provided in Table~\ref{tab:betas} and demonstrate that among each dropout sample, the typical UV slope is blue with a median $\beta$ that ranges from $-2.26$ to $-2.48$.
For the three lowest redshift samples, corresponding to F775W, F090W, and F115W dropouts, a sufficient number of galaxies are present for us to examine the UV luminosity dependence of their average UV slopes.
The F150W dropout sample contains just $24$ galaxies and only spans an \muv{} range of $-20.2$ to $-17.8$ making it challenging to constrain a trend with \muv{}.
Table~\ref{tab:betabins} gives properties of the bins within each dropout sample as well as the median $\beta$.
We now discuss the UV slopes of each sample individually.

\begin{table}
\begin{center}
\renewcommand{\arraystretch}{1.2}
\begin{tabular}{cccccc}
\toprule
Sample  & $z^{\rm med}_{\rm phot}$ & $d\beta/d\rm M_{\rm UV}$ & $\beta_0$ \\
\midrule
F775W Dropouts &   $5.86$ &  $-0.11\pm0.02$  & $-2.25\pm0.03$ \\
F090W Dropouts &  $7.28$ &   $-0.12\pm0.02$  & $-2.26\pm0.03$ \\
F115W Dropouts &  $9.41$ &  $-0.06\pm0.05$  & $-2.33\pm0.05$ \\
F150W Dropouts &  $12.02$ &  $-0.06\pm0.05^{\dagger}$  & $-2.42\pm0.13^{\dagger}$ \\
\bottomrule
\end{tabular}
\end{center}
\caption{Best-fit linear relations to UV slopes in bins of \muv{}. Uncertainties on the fit parameters are derived using a bootstrap Monte-Carlo method described in Section~\ref{sec:results}. The intercept ($\beta_0$) listed here is calculated at an \muv{}$=-19$.  $^{\dagger}$ As the F150W dropout sample is composed of only one \muv{} bin, we provide a fit assuming the slope is the same as our result for the F115W dropout sample.}
\label{tab:fits}
\end{table}

\subsubsection{F775W dropouts}
\label{sec:f775w}
We present the UV slopes found for the F775W dropout sample in Figure~\ref{fig:allslopes}.
These UV slopes have an inner 68th percentile range spanning in $\beta$ from $-1.7$ to $-2.7$, and comprise a median UV slope of $\beta=-2.26^{+0.03}_{-0.03}$.
This median value is broadly consistent with the UV slopes derived at $5.50<z<6.25$ from \citet{Nanayakkara2022} based on \jwst{} data from the GLASS \citep{Treu2022} program.
Figure~\ref{fig:sedexamples} provides photometry and best-fit SEDs for several galaxies in this sample across the range of observed UV slopes, illustrating the diversity of UV continua within.
Qualitatively, the UV slopes of the F775W dropout sample appear to shift toward bluer $\beta$ values at low UV luminosities, as been previously noted \citep[e.g.,][]{Finkelstein2012,Bouwens2014,Bhatawdekar2021}.
We explore this trend by assigning objects in our sample to bins in \muv{}, which are presented in Table~\ref{tab:betabins}.
We find that the most UV-luminous galaxies at this redshift (\muv{}$<-20$) display a median UV slope of $\beta=-2.12^{+0.06}_{-0.06}$, while the bin consisting of the faintest galaxies (\muv{}$>-17.5$) has a median UV slope of $\beta=-2.38^{+0.12}_{-0.13}$.
The number of galaxies at the redder end of the UV slope distribution ($\beta>-1.5$) appears constant as a function of \muv{}.
However, due to the increasing number of total galaxies, the fraction of $\beta>-1.5$ objects decreases toward low luminosities.

We quantify the observed trend between $\beta$ and \muv{} by fitting a linear relation to the median UV slopes for the luminosity bins of the form:
\begin{equation}
\label{eqn:1}
         \beta = \frac{d\beta}{d\rm M_{\rm UV}} \rm M_{\rm UV}^{-19}+\beta_0,
\end{equation}
where $\rm M_{\rm UV}^{-19}\equiv M_{\rm UV}+19$ and $\beta_0$ is the UV slope at an \muv{}$=-19$.
The best-fit parameters to this relation along with their uncertainties are presented in Table~\ref{tab:fits}.
We derive uncertainties on this best-fit relation using a bootstrap Monte-Carlo method, where we first construct a mock sample that matches the size and UV luminosity distribution of our observed sample, composed of objects selected randomly from the observed sample with replacement.
For each object in the mock sample, we perturb the rest-UV fluxes based on their uncertainties and measure the UV slope. 
We assign the resulting slopes to the same \muv{} bins described above, and fit the binned median $\beta$ using Equation~\ref{eqn:1}.
This process is repeated 1000 times, and we take note of the linear fit parameters for each iteration.
The uncertainties are then assigned as the range of the inner 68th percentile based on the resulting distribution.

This calculation reveals a significant correlation between $\beta$ and \muv{}.
We find a best-fit slope of $-0.11\pm0.02$ and normalization (defined at \muv{}$=-19$) of $-2.25\pm0.03$, which is inconsistent with no correlation between $\beta$ and \muv{} at $>5\sigma$.
Furthermore, a Spearman correlation test on the individual objects in the F775W dropout sample reveals a significant correlation, described by a correlation coefficient of $\rho_s=-0.25$, with an associated probability of being drawn from an uncorrelated distribution of $1.4\times10^{-6}$.
Using a sample of galaxies identified with \hst{}, \citet{Bouwens2014} derive a trend between $\beta$ and \muv{} with a slope of $-0.20\pm0.04$ for objects at \muv{}$=-22$ to $-17$, and a slope of $-0.08$ when just considering objects with luminosities fainter than \muv{}$>-18.8$.  
Together, these two estimates bracket the slope derived for our sample of F775W dropouts.
We find that the fit to our data is well described by a single component, such that we find a consistent trend when only considering the three most UV-luminous bins.
It is key to establish this trend among luminous systems, as the low-luminosity population is more susceptible to selection biases \citep[e.g.,][]{Rogers2013, Dunlop2012}.

\subsubsection{F090W dropouts}
\label{sec:f090w}
The total F090W dropout sample ($6.5\lesssim z \lesssim8.5$) is characterized by 
a median \muv{}$=-18.16$ and median UV slope of $\beta=-2.32^{+0.03}_{-0.02}$ (see Table~\ref{tab:betas}).
\citet{Nanayakkara2022} use \jwst{} data to derive UV slopes for a sample of galaxies at a consistent redshift to our F090W dropout sample.
For their sample, they find a median UV slope of $\beta=-2.30\pm0.26$, which is in excellent agreement with our results.
The UV slope distribution of the F090W dropout sample provided in Figure~\ref{fig:allslopes} appears qualitatively similar to that of the F775W dropouts, though key differences are present.
While the reddest sources at $z\lesssim6$ approach values of $\beta\simeq-1$, the F090W dropout sample contains no objects redder than $\beta>-1.4$.
However this does not represent a complete absence of moderately red ($\beta>-2.0$) sources. 
Two such examples are presented in Figure~\ref{fig:sedexamples}, which have UV slopes of $\beta=-1.98$ and $-1.44$.
Quantitatively, galaxies at $\beta>-1.5$ ($\beta>-1.2$) comprise a total of $9/656=1.3\%$ ($0/656=0\%$) of the F090W dropout sample, compared to $16/364=4.4\%$ ($11/364=3.0\%$) of the F775W comparison sample.

Galaxies with UV slopes bluer than $\beta<-2.5$ are present over a majority of the UV luminosity range sampled by the F090W dropouts.
The most luminous object in the sample with $\beta<-2.5$ is at \muv{}$=-20$, and such objects persist down to \muv{}$=-16.5$.
The presence of these blue sources is in spite of the varying uncertainty in UV slopes,
which typically are $0.03$, $0.20$, and $0.50$, at an \muv{}$=-21$, $-19$, and $-17$, respectively (see Figure~\ref{fig:allslopes}).
This implies significant width of the intrinsic UV slope distribution, hinting at variations in galaxy properties within the population at this epoch.
While the typical $\beta$ uncertainties grow large at low luminosities, many faint systems have UV slopes measured to much better precision, such as those falling within the deepest regions of the JADES footprint \citep{Eisenstein2023}.
As we note below, these low luminosity systems are crucial for our analysis into the population of extremely blue ($\beta\simeq -3$) galaxies.

We identify a clear trend between UV slope and \muv{} with low-luminosity galaxies displaying bluer UV continua on average.
We present the median UV slopes of these \muv{} bins in Table~\ref{tab:betabins}, and are displayed visually in Figure~\ref{fig:allslopes}.
Following the same procedure outlined above, we fit a linear relation to the median $\beta$, and find a trend with a slope of $-0.12\pm0.02$ and normalized to a $\beta=-2.26\pm0.03$ at \muv{}$=-19$.
This relation supports the trend of bluer UV slopes toward lower luminosities with a significance in excess of $5\sigma$.
To explore variations of this trend as a function of \muv{}, we fit a relation between each pair of bins, and in every case find a slope consistent with $d\beta/d\rm M_{\rm UV}=-0.12$.
This suggests that the observed trend is consistent with a single-component relation.
Previous studies have shown that faint galaxies with redder colors can be preferentially excluded due to photometric scatter, which biases the average UV slopes of the population to bluer values \citep[e.g.,][]{Dunlop2012, Rogers2013}.
To address this, we simulate the UV slope distribution analytically by scattering the source photometry by their uncertainties and passing the results through our selections.
We find  only a minimal bias at the faint end of our sample, which is smaller than the uncertainties in the median UV slope presented in Table~\ref{tab:betabins}.

\begin{figure}
    \centering
    \includegraphics[width=1.0\linewidth]{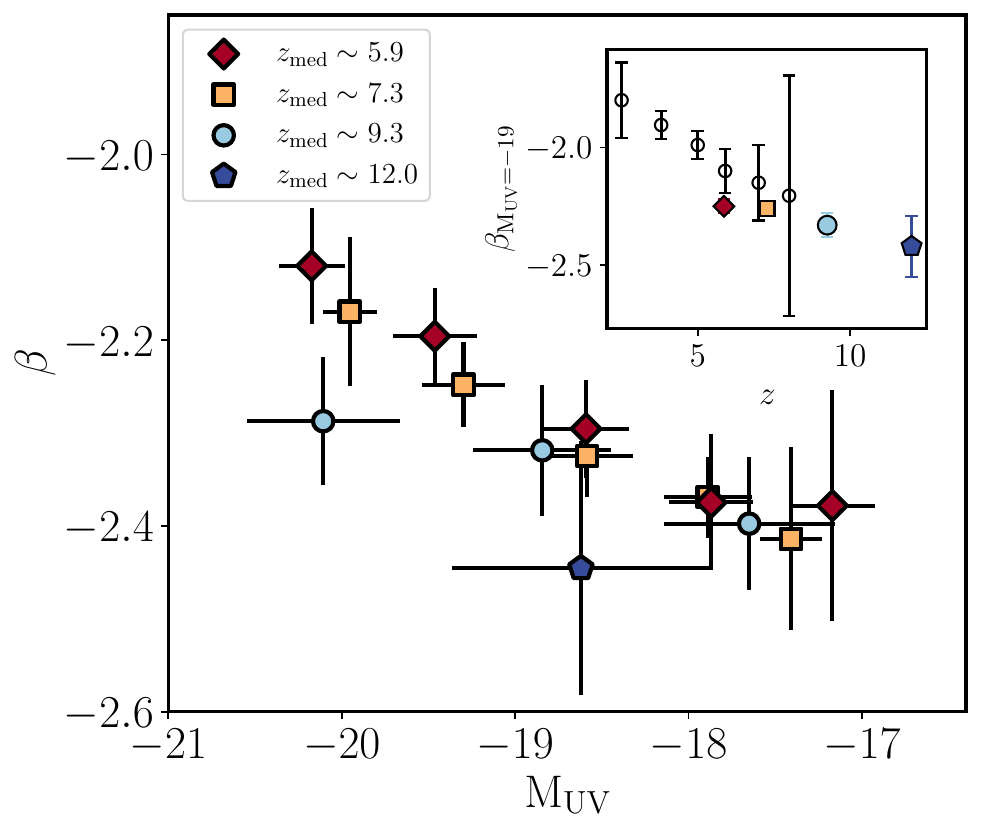}
    \caption{UV continuum slope, $\beta$, displayed as a function of $\rm M_{\rm UV}$ for the four dropout samples described in Section~\ref{sec:sample}. Each sample is divided into bins of \muv{}, with the exception of the F150W dropout sample median, which comprises the entire selection. For each sample explored in multiple bins of \muv{}, we observe a clear trend with $\beta$ such that fainter galaxies typically have bluer UV slopes. While this trend is clear throughout the different samples, the overall normalization decreases with increasing redshift. At fixed \muv{} the median UV slopes trend toward bluer values. The inset panel illustrates the median UV slope calculated at an \muv{}$=-19$ for our four dropout samples compared to the results from \citet{Bouwens2014} represented by small open circles.}
    \label{fig:muv-beta}
\end{figure}

\subsubsection{F115W dropouts}
\label{sec:f115w}
At $8.2\lesssim z \lesssim 11.3$, the F115W dropout sample ($z_{\rm med}=9.41$)
displays a median UV slope of $\beta=-2.35^{+0.04}_{-0.04}$.
The inner 68th percentile of the UV slope distribution spans a range of $\beta$ from $-2.8$ to $-1.9$, notably shifted toward bluer UV slopes when compared to the lower-redshift samples.
We present several examples of objects in our F115W dropout sample in Figure~\ref{fig:sedexamples}, where galaxies with UV slopes ranging from $\beta=-2.5$ to $-1.7$ are shown.
Among this sample, only five objects display UV slopes redder than $\beta=-1.5$, with the reddest galaxy displaying a UV slope of $\beta=-1.31$.
We find objects fitting this moderately red ($\beta>-1.5$) criteria at \muv{}=$-21$ as well as at \muv{}=$-17$.
However the majority of the reddest sources (4 out of 5) 
have absolute magnitudes placing them in the most luminous half of the sample (\muv{}$<-18.24$).

We divide the F115W dropout sample into bins of \muv{} to explore the trend between UV slope and UV luminosity.
Due to the number of F115W dropouts, we divide the sample into three bins, with median \muv{} of $-19.99^{+0.37}_{-0.55}$, $-18.93^{+0.31}_{-0.39}$, and $-17.98^{+0.42}_{-0.33}$.
The median UV slopes of each bin are provided in Table~\ref{tab:betabins}, with uncertainties derived using the same bootstrap Monte-Carlo method described above.
As with the lower-redshift samples, the median UV slopes of the F115W dropouts become increasingly blue at the faint end of the sample.
Specifically, we find a UV slope of $\beta=-2.28^{+0.06}_{-0.07}$ for the most UV-luminous bin at \muv{}$=-19.99$, trending to $\beta=-2.41^{+0.10}_{-0.10}$ at \muv{}$=-17.98$.
We fit a linear relation to these bins and find a best-fit relation of $\beta=(-0.06\pm0.05)(\rm M_{\rm UV}+19)+(-2.33\pm0.05)$.
Notably, the resulting best-fit relation is decidedly shallower than those of the lower-redshift F775W and F090W dropout samples.

A driving factor in this shallow slope is the blue color of the most UV-luminous galaxies in our sample.  As noted above, this stems mostly from the dearth of moderately reddened ($-2.0<\beta<-1.5$) objects at high UV luminosities in the F115W dropout sample. 
Excluding the bin at \muv{}$\simeq-20$ from the linear fit reveals a much steeper trend with $\beta$, with a relation of $\beta=-0.09\times (\rm M_{\rm UV}+19)-2.33$.
Given this relation, we would expect the median UV slope of the most luminous bin to be $\beta=-2.24$, which is within the uncertainty of the observed median.
One possible cause for this difference among bright galaxies results from the relatively small sample size of the brightest bin.
In contrast, the intermediate UV luminosity bin has over double the sample size, which contains 18 galaxies within the UV slope range $-2.0<\beta<-1.5$.
If a consistent fraction were present among the brightest bin, we would expect eight such galaxies in the most UV luminous bin.

\subsubsection{F150W dropouts}
\label{sec:f150w}
The F150W dropout sample includes the highest-redshift objects analyzed in this work, comprising 24 galaxies spanning $11<z_{\rm phot}<14$.
A majority (20) of the objects satisfying these selection criteria have already been presented in \citet{Hainline2023}, and the remaining four objects (that are very faint) will be discussed further in Whitler et al. (in prep).
Due to the small number of galaxies falling into this selection, and relatively narrow range of \muv{} spanned by the sample, we are unable to construct multiple bins of \muv{}.
Thus, we explore the F150W dropout sample as a whole, finding a median \muv{} of $-18.61^{+0.81}_{-0.91}$ along with a median redshift of $z_{\rm phot}=12.02$ (see Table~\ref{tab:betas}).
At the redshifts probed by this sample, the rest-optical emission from galaxies has redshifted beyond wavelengths probed by \jwst/NIRCam, making UV slopes one of the few remaining observational probes of galaxy properties.
For this F150W dropout sample, we find a median UV slope of $\beta=-2.48^{+0.12}_{-0.14}$.
We present several examples of objects selected as F150W dropouts in Figure~\ref{fig:sedexamples}, illustrating the variety of $\beta$ observed in this sample despite the relatively small sample size.

Strikingly, the F150W dropout sample constitutes an epoch where we observe the disappearance of a significant portion of the sample with $\beta\gtrsim-2.0$.
We test our ability to identify such sources by artificially scattering the photometry of our sample by their uncertainties such that they fall into this moderately-red regime.
This investigation reveals that we do recover these galaxies, suggesting that the lack of a $\beta\gtrsim-2.0$ population is not driven by selection effects.
While we have described some differences in the abundance of such sources among our lower-redshift samples, the F150W dropouts are markedly distinct in this regime.
Such systems comprise $22.8$, $17.5$, and $19.4$ per cent of objects among our samples at $z_{\rm phot}=5.86$, $7.28$, and $9.41$, respectively.
Adopting a similar fraction for the F150W dropout sample, we may expect $\sim5$ of these objects to have been selected in our sample.
While only one such object has found its way into the selection, larger samples of galaxies in this redshift range may reveal better consistency with the lower redshift samples, or confirm the dearth of moderately red galaxies at the earliest epochs.

As the number of F150W objects in our sample is small, we are not able to place meaningful constraints on the \muv{} dependence of the UV slopes.
However, we explore the trend that may exist if we make an assumption on the slope of the relation.
Specifically, if we adopt the slope derived from the F115W dropouts (e.g., $d\beta/d\rm M_{\rm UV}=-0.06$), the normalization provided by the median F150W dropout slope yields a relation of $\beta=-0.06\rm (M_{\rm UV}+19)-2.42$.
We provide the parameters of this fit in Table~\ref{tab:fits}, but note that it was derived assuming a slope fixed to that of the F115W dropouts.
If instead we adopt the steeper slope found for the F090W dropouts (e.g., $d\beta/d\rm M_{\rm UV}=-0.12$), we achieve a relation of $\beta=-0.12\rm (M_{\rm UV}+19)-2.43$.
These two relations make substantially different predictions when extrapolated to values of \muv{} beyond the extent spanned by the sample.
Specifically, at an \muv{} of $-20$, the steeper of these relations predicts a $\beta$ of $-2.31$, which would be consistent with the UV-bright F115W dropouts, while the shallower of these relations would yield a $\beta=-2.37$.
Larger samples that can expand the dynamic range of \muv{} probed in this epoch will unveil any UV-luminosity dependence of $\beta$ among this population.

\begin{figure*}
    \centering
    \includegraphics[width=1.0\linewidth]{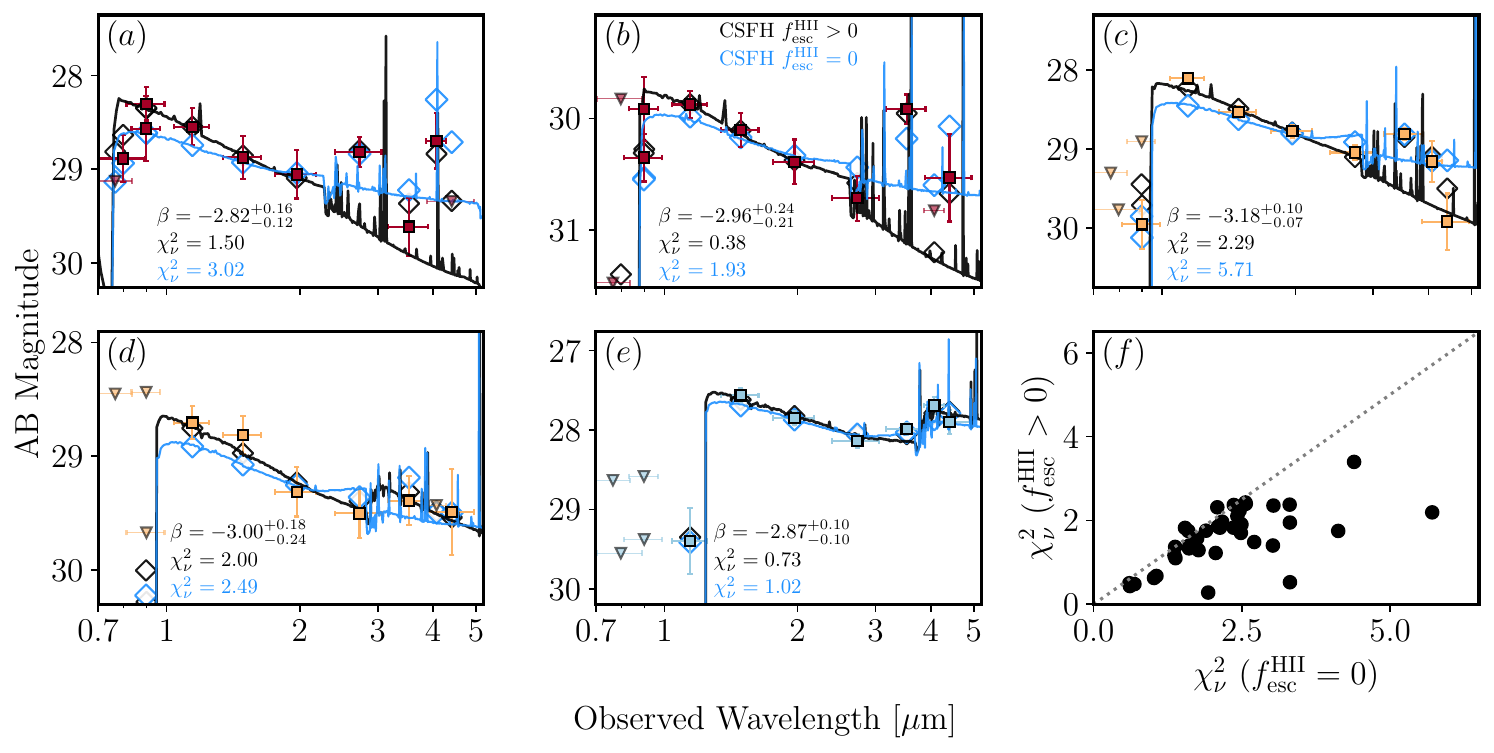}
    \caption{Demonstration of improved fit quality when templates with \fesc{}$>0$ are used to model the photometry of our extremely blue sample. Panels (a)-(e) show best-fit SED models with \fesc{}$=0$ are displayed in blue, while the models allowing for high \fesc{} are shown in black. Photometry are color-coded based on their dropout sample as defined in Figure~\ref{fig:sample}. We list the reduced $\chi^2$ for both of these model fits in each panel. Panel (f) directly compares the reduced $\chi^2$ between the two model fits for the entire extremely blue sample. In nearly all cases, allowing for high values of \fesc{} yields a much better fit, indicated by the lower $\chi^2$.}
    \label{fig:fesccompare}
\end{figure*}

\begin{figure*}
    \centering
    \includegraphics[width=1.0\linewidth]{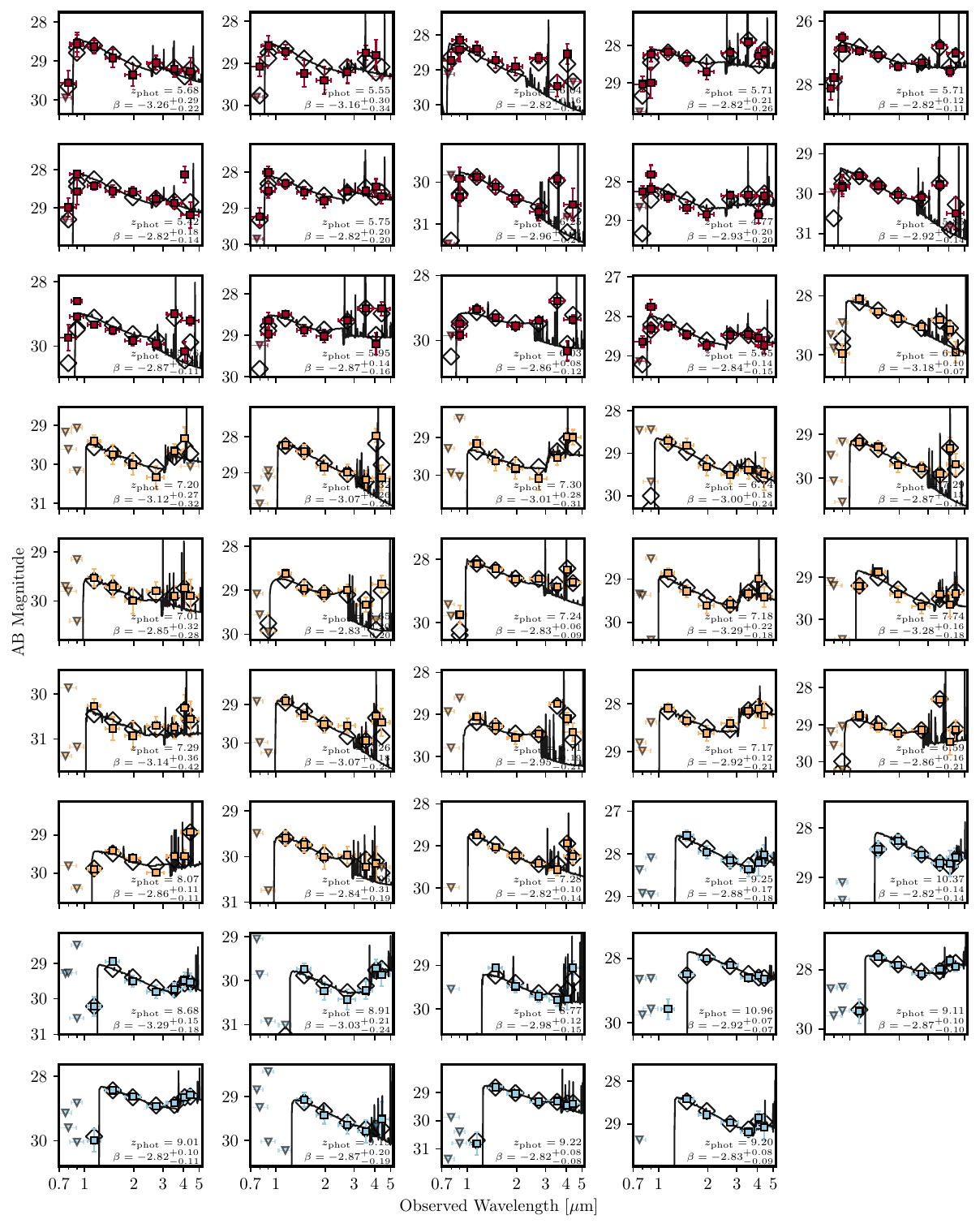}
    \caption{Best-fit SEDs for galaxies with extremely blue UV slopes discussed in Section~\ref{sec:vblueselection}. In each panel we display the photometric redshift and UV slope along with their uncertainties. The observed fluxes are displayed according to the sample in which the objects derive from, such that F775W, F090W, and F115W dropouts are displayed as red diamonds, orange squares, and blue circles, respectively.}
    \label{fig:robustSED}
\end{figure*}

\subsection{Redshift evolution of typical UV slopes from $z=6-12$}
\label{sec:zevol}

The typical UV slopes derived for the four dropout samples demonstrate a key aspect of evolution in galaxy properties at $z\gtrsim 6$.
The redshift evolution is apparent in Figure~\ref{fig:muv-beta} where the median UV slopes become more blue toward higher redshift at fixed \muv{}.
At an \muv{}$=-19$ (which is well sampled by each of our dropout samples), we find a shift in the typical UV slopes from $\beta=-2.25$ at $z\sim6$ to a $\beta=-2.42$ at $z\sim12$.
The results for each of our samples are presented in the inset panel of Figure~\ref{fig:muv-beta}.
We quantify the strength of this trend by fitting a linear relation to these medians, and find that they imply a redshift evolution with $\beta$ at a rate of $d\beta/dz=-0.030^{+0.024}_{-0.029}$.

This derived evolution provides an extension upon previous results finding a similar trend among lower-redshift samples \citep[e.g.,][]{Finkelstein2012, Bouwens2012, Bouwens2014}.
Up to $z\sim6$, \citet{Bouwens2014} derive a UV slope evolution of $d\beta/dz=-0.10\pm0.06$, which is broadly consistent with the evolution derived from our \jwst{} sample.
The median UV slope from our $z\sim6$ (corresponding to F775W dropouts) sample lies roughly $1\sigma$ below the \citet{Bouwens2014} measurement at the same redshift.
This slight discrepancy is one element causing our estimate of the redshift evolution be slower than what has been found by \citet{Bouwens2014}.
For redshifts of $6\lesssim z \lesssim 8$ (i.e., where the two samples overlap), we find good consistency between the median UV slopes.
At yet higher redshift, our results are in agreement with the typical UV slopes obtained when extrapolating the \citet{Bouwens2014} trend to earlier times.
Using just these objects in our sample ($z>8$) we measure an evolution of $d\beta/dz=-0.07^{+0.05}_{-0.06}$, consistent with the trend derived at lower redshift in \citet{Bouwens2014}.

The JADES results suggest the evolution in UV colors proceeds differently at different UV luminosities.
We find significantly faster evolution toward bluer colors at  \muv{}$=-20$ than  at \muv{}$=-18$, amounting to a $d\beta/dz=-0.038\pm0.017$ and $d\beta/dz=-0.008\pm0.017$, respectively. 
As is clear in Figure~\ref{fig:muv-beta}, the lowest luminosity galaxies in our sample show minimal evolution between $z\simeq 6$ and $z\simeq 9$.
This may reflect a larger fraction of the lower luminosity population approaching the intrinsic values expected for normal stellar populations in the absence of dust by $z\simeq 6$, thereby saturating the blueward evolution  \citep{Cullen2017}. 
In contrast, the most luminous galaxies in our sample show marked evolution between $z\simeq 7$ and $z\simeq 9$, with very blue 
slopes  ($\beta=-2.3$) becoming typical for \muv{}$=-20$ at $z\simeq 9$. This may suggest rapid evolution in the dust content of the 
most UV luminous galaxies, an effect that has been noted as a possibility to address the high number of UV-luminous systems appearing at $z>10$ \citep[e.g.,][]{Ferrara2023, Mason2023}.
In addition, larger fractions of a population approaching this intrinsic limit
ensures that several prerequisites (e.g., no dust) are in place that allow for extreme UV slopes (e.g., $\beta\simeq-3$) to exist in greater abundance.
We describe such objects within our sample in detail in the following section.

The UV slope evolution described in this section demonstrates that galaxies at $z\gtrsim9$ are unmistakably blue on average.
At $z\sim9.3$ and $z\sim12.0$ we find median UV slopes of $\beta=-2.33^{+0.09}_{-0.08}$ and $\beta=-2.48^{+0.12}_{-0.14}$, implying evolution continues to the highest redshifts.
The ubiquity of blue colors in galaxies at these epochs is also present within semi-analytic galaxy evolution models \citep[e.g.,][]{Mirocha2023, Yung2023}, driven by a range of physical effects (e.g., dust attenuation, stochasticity of star formation).
\citet{Mirocha2023} utilized these theoretical models to explore the processes that drive the apparent over-abundance of UV-bright sources at early times.
In doing so, they find that galaxies with UV slopes of $\beta\simeq -2.0$ to $-2.5$ are typical at $z>8$, in agreement with our results.
However, the physical processes imposed in the models of \citet{Mirocha2023} may not be the only way to reproduce observed galaxy abundances.
Additional insight into the evolving UV slope distribution, such as the absence of moderately-red sources at $z\sim12$ noted in this work, will provide significant leverage to inform future models of galaxy evolution.

\begin{figure}
    \centering
    \includegraphics[width=1.0\linewidth]{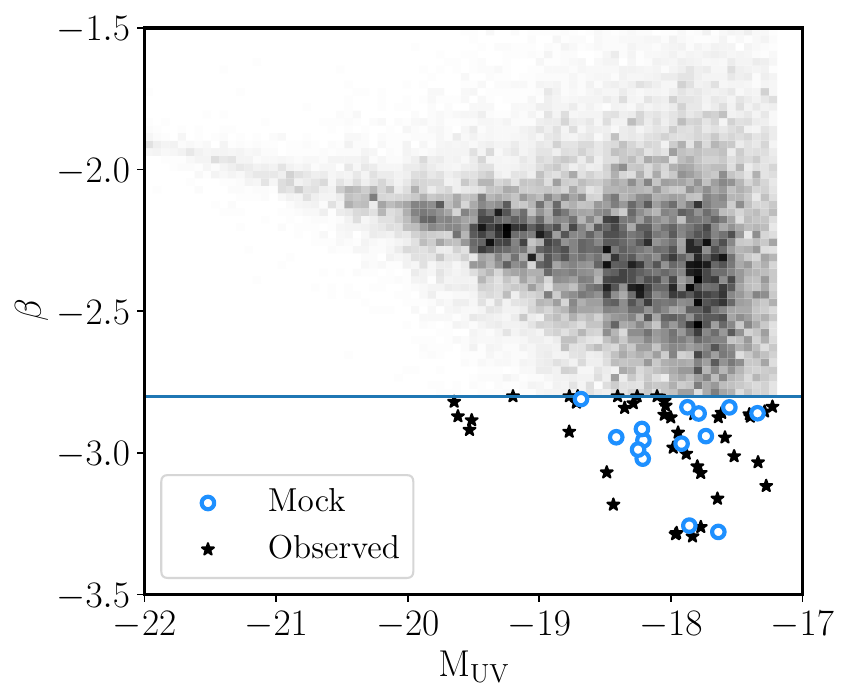}
    \caption{Simulated UV slope distribution constructed from the measured intrinsic relation between $\beta$ and \muv{} with simulated scatter derived from the photometric flux uncertainties. The mock sample is designed to have demographics matching that of our observed sample as described in Section~\ref{sec:sample}. We indicate the $\beta=-2.8$ threshold for considering an object as extremely blue with a horizontal line.  Despite a fixed intrinsic width of the UV slope distribution, the decreasing S/N measured for fainter systems results in a significant increase in the observed $\beta$ scatter. We impose our criteria for robust extremely blue objects described in Section~\ref{sec:vblueselection}, and display the resulting sample as blue circles. For comparison, we plot our observed sample of robust extremely blue objects as black stars. It is clear that while photometric scatter is capable of producing galaxies falling into our robust selection, it appears to not be able to explain the observed sample, both in number and in UV luminosity range.}
    \label{fig:mock}
\end{figure}

\begin{figure*}
    \centering
    \includegraphics[width=1.0\linewidth]{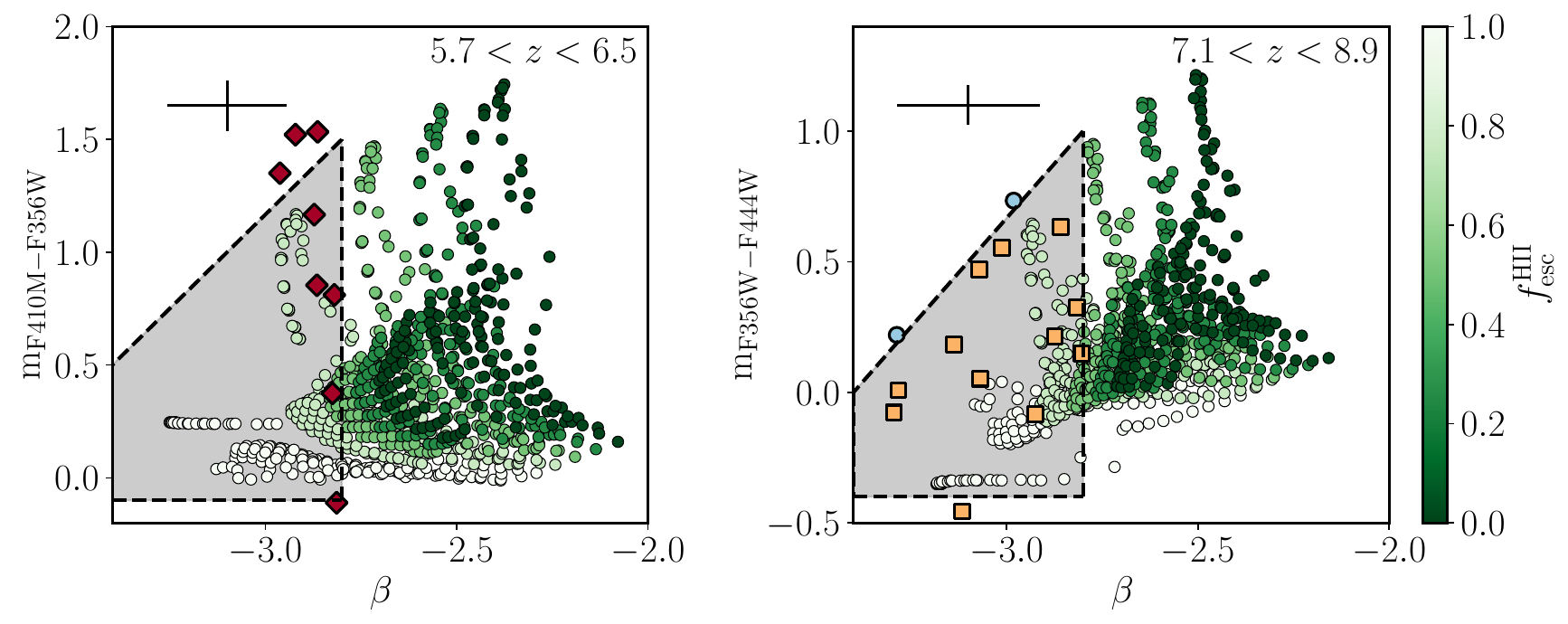}
    \caption{Inferred emission line strengths of the extremely blue sample. The left panel shows the selection at redshifts of $5.7<z<6.5$, while the right panel is for $7.1<z<8.9$. In each panel, we display colors and UV slopes calculated from model templates for five values of \fesc{}, and over a grid of metallicity ($\log(Z/Z_{\odot})\in[-2,0]$) and age ($\log(\rm Age/yr)\in[6,9]$) as the green points. Galaxies within our sample selected as F775W, F090W, and F115W dropouts are displayed as red diamonds, orange squares, and blue circles, respectively. The shaded region enclosed by black dashed lines indicate our selection of robust objects with extremely blue UV slopes and relatively weak lines. We provide the median error bars in the top left of each panel.}
    \label{fig:blueselection}
\end{figure*}

\begin{figure*}
    \centering
    \includegraphics[width=1.0\linewidth]{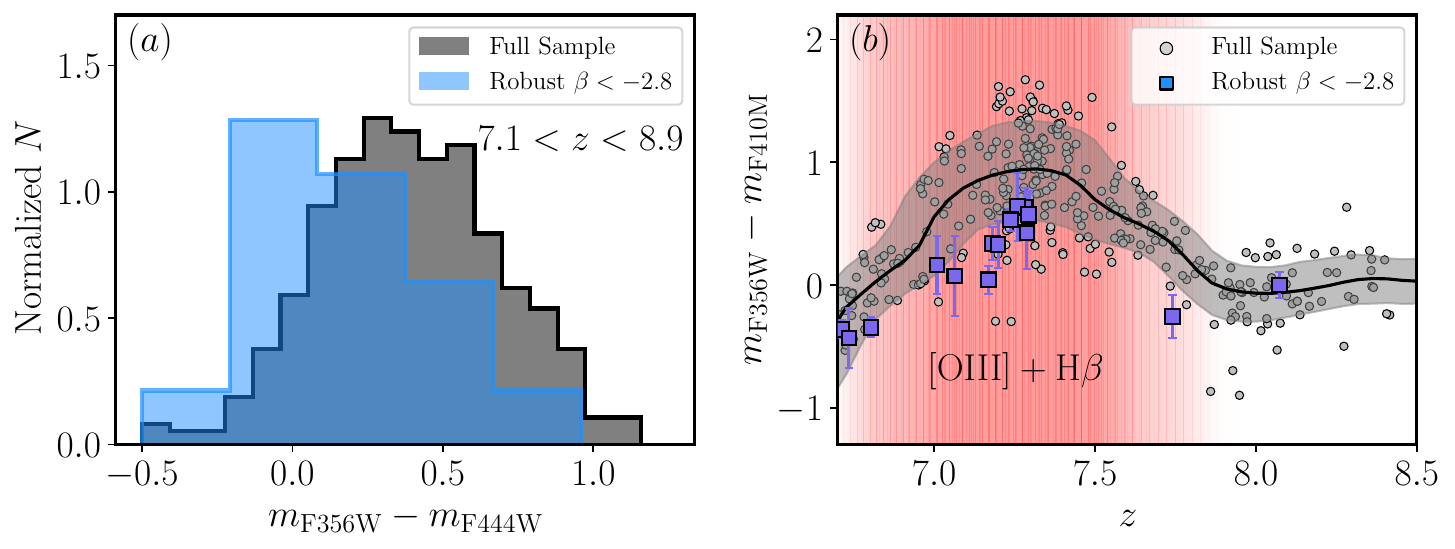}
    \caption{Comparison of color excesses of the extremely blue galaxies compared to the full dropout samples. (a): F356W-F444W colors for galaxies with photometric redshifts at $7.1<z<8.9$, where this LW color is sensitive to the $\rm [OIII]+H\beta$ EW. We present colors for the full sample in the grey histogram, which has a median value of $0.39$ mag. In comparison, we show colors from galaxies in our extremely blue sample in the blue histogram with a median value of $0.19$ mag implying weaker emission lines compared to the full sample. (b): F410M narrow-band excess as a function of redshift for the full sample (grey points) and the extremely blue objects (blue points). At $7.0\lesssim z \lesssim 7.5$, $\rm [OIII]+H\beta$ contributes to the F410M flux, resulting in a color excess (red shaded region). These colors imply that the extremely blue galaxies have weaker lines at fixed redshift than the full sample on average.}
    \label{fig:weakexcess}
\end{figure*}

\begin{figure*}
    \centering
    \includegraphics[width=1.0\linewidth]{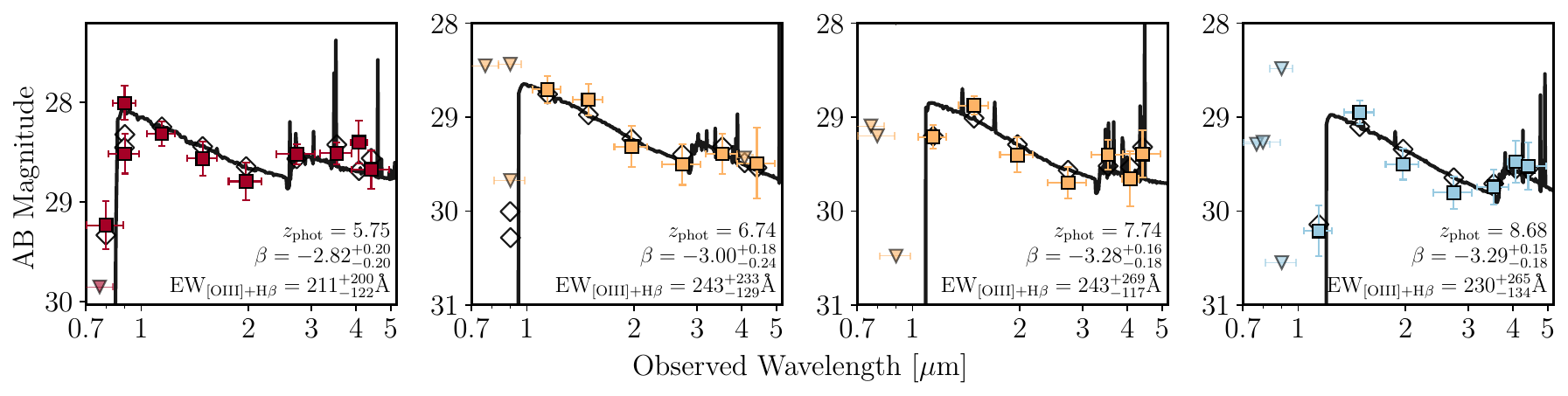}
    \caption{Examples of galaxies in our extremely blue sample that display no significant photometric excesses, implying weak rest-optical emission lines. In each panel, we present the UV slope measured from the photometry, and photometric redshift and $\rm [OIII]+H\beta$ EW inferred from the best-fit BEAGLE model. The photometric points are color-coded based on their dropout sample as defined in Figure~\ref{fig:sample}. The weak lines and extremely blue UV slopes imply low amounts of nebular emission, indicative of high \fesc{} within these galaxies.}
    \label{fig:weaklines}
\end{figure*}

\begin{figure}
    \centering
    \includegraphics[width=1.0\linewidth]{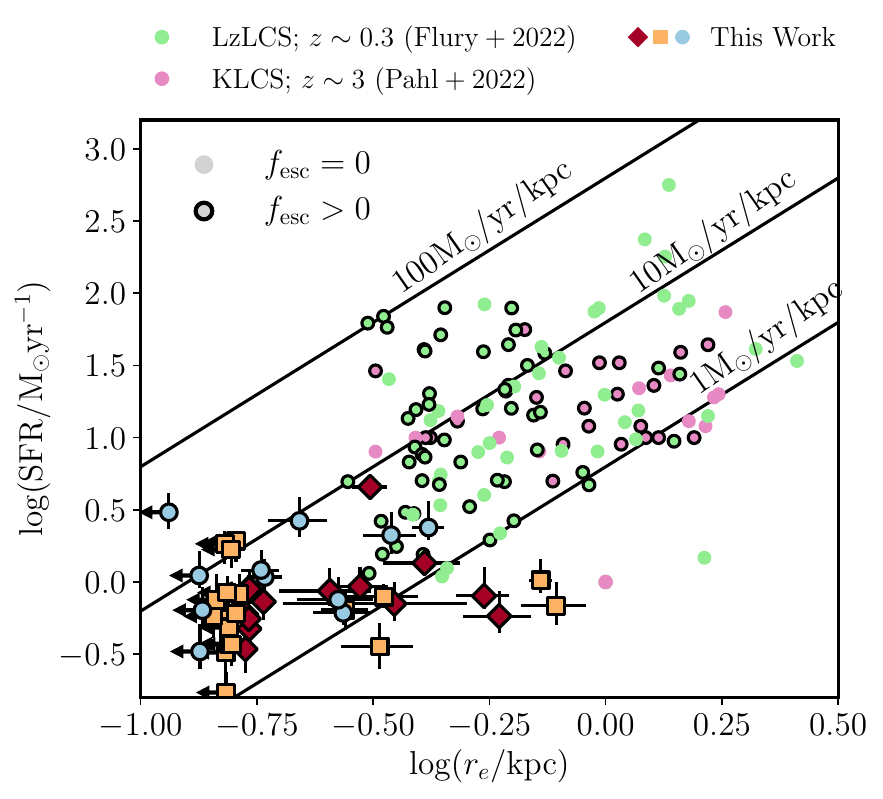}
    \caption{Star-formation rates derived from the best-fit BEAGLE models as a function of effective radius measured from the NIRCam mosaics. The radii presented here are measured for each object in the filter corresponding to the rest-frame 1500\angstrom{}. We display and label lines of constant SFR surface density ($\Sigma_{\rm SFR}$). Points in our sample are displayed using the same scheme as in previous figures. For comparison, we display objects from the LzLCS \citep{Flury2022a} comprising a sample of $z\sim0$ galaxies, in addition to a sample of galaxies at $z\sim3$ from \citet{Pahl2021}. Galaxies in the low-redshift comparison samples with confirmed $f_{\rm esc}>0$ are outlined in black. The sources in our sample typically have much smaller SFRs and are more compact compared to the low-redshift objects. }
    \label{fig:sfrdensity}
\end{figure}

%
%
\section{Galaxies with very blue UV slopes in JADES}
\label{sec:discussion}

We now investigate the bluest galaxies in the JADES 
database at $z\gtrsim 6$ with UV slopes approaching $\beta \simeq -3$. Interest in this population is driven by their potential for indicating LyC photon leakage or exotic stellar populations \citep[e.g.,][]{Bouwens2010, Ono2010, Robertson2010, Zackrisson2013, Jiang2020, Topping2022,Cullen2022}. In 
this section, we first identify our most stringent 
population of very blue galaxies (\S4.1) before 
exploring their properties (\S4.2) and discussing the 
potential impact of bursty SFHs on the observed colors (\S4.3).

\subsection{A population of extremely blue UV objects at $z\gtrsim6$}
\label{sec:vblueselection}

As is clear from Figure~\ref{fig:allslopes}, a significant subset of the 
the $1276$ galaxies in our dropout samples have observed UV slopes nearing $\beta=-3$. Our first goal in this section is to identify those galaxies which 
are most likely to have very blue UV slopes. We 
find that $14\%$ (184) of objects across the full sample have UV slopes of $\beta<-2.8$. In this regime, 
standard population synthesis models struggle to 
explain the colors with ionization-bounded ISM conditions. However interpretation of such blue 
galaxies has long been impacted by photometric uncertainties, as the error on $\beta$ is often large enough such that the blue UV slopes can be the result of anomalous scattering.

We seek to identify the subset of very blue ($\beta<-2.8$) sources that are most robust against 
concerns of photometric scatter. To achieve this 
goal, we first require that the uncertainty on the UV slope is small enough ($\sigma_{\beta}<0.3$) to minimize inclusion of very noisy sources likely to have 
significant scatter in their $\beta$ measurements
(see also \citealt{Cullen2022} for an alternative approach). This reduces the very blue sample to $62$ objects.
Following \citet{Topping2022}, we fit this sample with density-bounded templates \citep{Plat2019} that allow UV slopes to approach the very blue values ($\beta\simeq -3.2$) observed in this sample. 
These models allow for nebular emission to be self-consistently computed while allowing for the escape of ionizing photons in the density-bounded regime. 
The density-bounded models generally result in good fits to the observed SEDs, making significant improvements with respect to the standard ionization-bounded models.  This is clearly seen in the SEDs shown in  Figure~\ref{fig:fesccompare}, with  
$\chi^2_{\nu}$ improving from  $0.5-12.1$ (median $2.2$) in the \fesc{}$=0$ models to  $0.3-9.7$ (median $1.8$) when \fesc{}$>0$ models are allowed (Figure~\ref{fig:fesccompare}f).

In some cases, we see that the observed blue UV slopes may be significantly influenced by one photometric filter that is  offset from the SED. 
For example, if the observed flux in the filter just redward of the Ly$\alpha$ break is well in excess of the 
SED (owing to photometric scatter), it can lead to an anomalously blue UV slope. 
We conservatively excise such objects from the sample. 
To identify these sources, we select galaxies 
where the observed flux density in one or more of the rest-UV 
filters is significantly offset from the best-fit density-bounded models (i.e., by more than the photometric uncertainty in that filter). 
With this cut, our extremely blue sample reduces to $44$ galaxies, comprising $3.4\%$ of our total sample.
Of these $44$ galaxies,  $14$, $19$, and $11$ objects are in the F775W, F090W, and F115W selections, respectively.   We present the SEDs for each object in this sample in Figure~\ref{fig:robustSED}.
This extremely blue sample spans a range of redshifts from $z_{\rm phot}=5.42-10.96$ ($z_{\rm med}=7.18$) and UV luminosities corresponding to \muv{}$=-19.7$ to $-17.2$ (median \muv{}$=-18.0$).
The UV slopes span from $\beta=-2.81$ to $-3.26$ ($\beta_{\rm med}=-2.95$), with typical uncertainties of $0.23$, much less than the typical uncertainties found for galaxies in our full sample at these \muv{} (e.g., $0.43$ at \muv{}$=-18$; Section~\ref{sec:UVslopes}).  The density-bounded models require 
fairly large escape fractions to match the observed SEDs (median \fesc{}$=0.51^{+0.08}_{-0.09}$), with the highest \fesc{} in the sample reaching $0.86$.
A small fraction of this sample (6/44) have inferred \fesc{} of less than $20\%$.
This small number of objects with lower \fesc{} typically have UV slopes near the edge of the selection (i.e., $\beta\simeq-2.8$).

Having identified a small sample of galaxies with very blue UV colors, we now more closely investigate the 
impact of photometric scatter on our measured colors. We are in particular interested in determining if photometric uncertainties are expected to scatter significantly more sources into 
the extremely blue regime ($\beta <-2.8$) than are observed.  
We quantify the number of sources expected to scatter to very blue colors using a simple set of simulations that mock up galaxies 
with the same demographics as those found for this analysis (see Section~\ref{sec:sample}).
This mock sample is constructed independently for each of the dropout samples described above.
We randomly select a UV luminosity from the \muv{} distributions presented in Section~\ref{sec:sample}.
From this UV luminosity, we assign a UV slope based on the relations in Section~\ref{sec:UVslopes}, to which we apply a scatter of $0.2$, consistent with the spread of $\beta$ found for objects at the 
bright end of our sample where photometric scatter is minimal (and 
the observed scatter is assumed to be intrinsic). 
Based on the selected \muv{} and assigned $\beta$, we derive mock  photometry in filters sampling the rest-frame UV.
We next assign flux uncertainties to the mock SEDs. The uncertainties are randomly selected from our observed sample for objects that are matched in \muv{}, naturally approximating 
the varying observing depths across the JADES mosaics \citep{Eisenstein2023}.
We perturb the mock photometry by their assigned uncertainties, and measure the UV slope as described in Section~\ref{sec:properties}.
This process is repeated until the mock sample contains the same number of objects as each observed dropout sample. 

We present a typical realization of these simulations in Figure~\ref{fig:mock}.
As expected, the observed scatter of UV slopes for this mock sample increases greatly at low UV luminosities.
By passing the mock samples through our criteria for robust extremely blue sources, we can estimate the number of objects expected to exist due to scatter.
At all UV luminosities, these simulations under-predict the number of robust extremely blue sources.
The robust extremely blue objects comprise $2.0\%$, $3.2\%$, and $4.5\%$ of the observed sample at $-20\leq \rm M_{\rm UV}<-19$, $-19\leq \rm M_{\rm UV}<-18$, and $-18\leq \rm M_{\rm UV}<-17$, respectively, while the robust objects in these simulations yield only $0\%$, $1.6\%$, and $2.6\%$, in the same UV luminosity bins.
In total, scattering these galaxies to blue UV slopes yields less than half of the number of robust objects of our observed sample ($32\%$ on average), implying that photometric scatter alone cannot explain the observed sample of extremely blue objects with our adopted assumptions on intrinsic scatter.
This analysis has primarily focused on the rest-UV measurements, while in the following section we determine if the extremely blue UV slopes are supported by the rest-optical SEDs.

\subsection{Physical properties of galaxies with very blue UV slopes}
\label{sec:fesc}

In the previous section, we assembled a sample of 
$z\gtrsim 6$ galaxies in the JADES footprint with very blue colors ($\beta < -2.8$). We demonstrated that density-bounded photoionization models \citep{Plat2019} can successfully reproduce the SEDs with a typical escape fraction of \fesc{}$=0.51$. Here we begin to investigate the physical properties of the sources. We first explore the rest-optical SEDs, with particular attention on the emission line 
strengths. We then investigate whether the properties implied by the SEDs are distinct from the general population and compare the star formation rate surface densities to known population of LyC leakers.

If the very blue galaxies are leaking a large fraction of their ionizing radiation, we would expect their rest-optical emission lines to be significantly weakened  
\citep[e.g.,][]{Zackrisson2013}. The influence of these emission lines is readily seen in flux excesses in medium and broadband NIRCam filters \citep[e.g.][]{Endsley2023}.
Using the JADES filters, the LW colors are sensitive to the $\rm [OIII]+H\beta$ EWs within two redshift windows.
Within $5.9<z<6.6$ ($7.1<z<8.9$), the $\rm [OIII]+H\beta$ lines fall in the F356W (F444W) filter, while the F410M (F356W) filter is sensitive to the continuum only.

We compare the observed UV slopes and LW colors of the extremely blue sample to models of density-bounded HII regions.
For this comparison, we construct a grid of models using templates from \citet{Plat2019} for values of \fesc{} from $0$ to $1$ in steps of $0.25$, in addition to a broad range of ages (1-300Myr) and metallicities ($\log(\rm Z/Z_{\odot})=-2$ to $0$) in steps of $0.1$ dex.
We present the colors and UV slopes derived from these models in Figure~\ref{fig:blueselection}.
These models illustrate two clear characteristics; higher values of \fesc{} allow models to achieve bluer UV slopes, and at bluer UV slopes the color excesses (i.e., emission line EWs) are weaker \citep[e.g.,][]{Zackrisson2013, Topping2022}.
We define the following regions of the color-$\beta$ space that encompass the model predictions at $\beta<-2.8$ and corresponding to high \fesc{} ($>50\%$).
\begin{equation}
\begin{cases}
\beta<-2.8 \\
F410M-F356W > -0.1 \\
F410M-F356W < 1.67\times\beta + 6.16 \\
\end{cases}
\end{equation}
for galaxies within the redshift range $5.7<z<6.5$, and 
\begin{equation}
\begin{cases}
\beta<-2.8 \\
F356W-F444W > -0.4 \\
F356W-F444W < 1.67\times \beta + 5.67 \\
\end{cases}
\end{equation}
for galaxies at $7.1<z<8.9$. 
These two regions are highlighted in Figure~\ref{fig:blueselection}.

The colors and UV slopes of our extremely blue sample are also compared to these regions in Figure~\ref{fig:blueselection}.
In total, 22 ($50\%$) galaxies from our extremely blue sample fall within the redshift ranges described above.
This Figure illustrates that nearly all galaxies within this redshift-restricted subsample (16/22) fall within the $\beta$-color space informed by the models.
The remaining six objects are slightly outside the boundary of these two regions, however they are consistent within the uncertainties.
It is clear that the bluest subset ($\beta<-3.0$) of our $\beta<-2.8$ sample possesses the weakest rest-optical color excesses. 
We find a median $m_{\rm F356W}-m_{\rm F444W}$ color for this bluest subset of $0.11$ mag, while typical colors of objects with $-2.8<\beta<-3.0$ are elevated to $0.27$ mag.
These colors imply $\rm [OIII]+H\beta$ EWs of $221$\angstrom{} and $317$\angstrom{} for the $\beta<-3.0$ and $-2.8<\beta<-3.0$ subsamples, respectively.
The comparatively weak lines inferred for these sources provide a key self-consistent check on expectations from models with high \fesc{}.

We explore the rest-optical emission lines further in Figure~\ref{fig:weakexcess}. Here we compare the color excesses of our extremely blue objects to a representative sample of galaxies pulled from the selection in Section~\ref{sec:sample}.
First concentrating on the subsample of objects $7.1<z<8.9$, we find a median F356W-F444W color for the extremely blue objects of $0.19$ mag, which is systematically lower than the $0.39$ mag found for the full sample.
This difference is reflected in the $\rm [OIII]+H\beta$ EWs; the extremely blue objects are characterized by a median EW of $256$\angstrom{}, which are systematically weaker than the $340-780$\angstrom{} expected for typical galaxies in this epoch \citep{Endsley2023}.
It is conceivable that the extremely blue galaxies lie at redshifts where the observed-frame wavelengths of $\rm [OIII]+H\beta$ lines correspond to low-throughput regions of the filter sensitivity curves.
In turn, this would cause the emission lines to be less effective at producing color excesses.
To explore this possibility, we display the F356W-F410M colors as a function of photometric redshift in Figure~\ref{fig:weakexcess}b.
For this test, we use the narrower F410M filter, as it is more sensitive to the presence of emission lines, however over a smaller range of redshifts.
Throughout the redshift range where $\rm [OIII]+H\beta$ contributes to the F410M flux, the extremely blue sample yields LW colors implying emission lines that are weaker than the full sample on average.
The median F356W-F410M color of the full sample reaches its maximum at a redshift of $z\simeq7.3$ and has a value of $1$ mag.
Extremely blue objects at the same redshift are all below this value, and have a maximum color of $0.6$ mag, and primarily fall into the range of $0.3-0.5$ mag.
This provides additional evidence that the extremely blue sample displays systematically weaker emission lines than the overall galaxy population, in agreement with their leaking of significant ionizing radiation.

Figure~\ref{fig:weaklines} presents SEDs for a subset of our extremely blue sample that appear to have weak emission lines.
Each object displayed here lies at $z<8.9$, such that the $\rm [OIII]+H\beta$ lines fall within the NIRCam filters.
While a visual inspection of these SEDs already indicates the absence of strong emission lines, the full SED modelling of their photometry confirms that the emission lines are very weak, with BEAGLE-derived EWs of only $200-250$\angstrom{}.
In combination, extremely blue UV slopes and weak rest-optical emission lines provide a compelling case for high \fesc{} among these galaxies.
We note that weak emission lines are not uniquely caused by high \fesc{}, however the alternative solutions such as extremely low metallicities or complex SFHs (see \citealt{Endsley2023}) yield different predictions for the overall SED.
For example, extremely low metallicities would result in a strong nebular continuum component of the emission, making UV slopes of $\beta<-2.8$ impossible. In contrast, 
complex SFHs may be able to reproduce the observed SEDs, which we discuss in detail in the following section.

We next explore the SED-based properties of this candidate population of high-\fesc{} systems. Previous analyses have argued using both observational \citep[e.g.,][]{Heckman2011} and theoretical work \citep[e.g.,][]{Rosdahl2022}  that intense feedback can drive high escape fractions, with such feedback being 
particularly significant among young, high-sSFR galaxies.
These characteristics are common among our extremely blue sample which is typically young (median CSFH age $14.8_{-5.8}^{+9.8}$Myr), with average sSFRs of $79~\rm Gyr^{-1}$.
This sSFR is nearly a factor of two greater than the median found for the full sample ($44~\rm Gyr^{-1}$).
In a similar manner, much debate has centered on whether higher escape fractions are more prevalent among lower or higher stellar mass galaxies \citep[e.g.,][]{SaldanaLopez2022, Naidu2020, Ma2020} due to the interplay of star formation, galaxy geometry, and dust attenuation (see e.g., \citealt{Pahl2023} for a discussion).
We find that all of the objects in our extremely blue sample have masses in the range $\log(\rm M/M_{\odot})=7.0-8.5$, with a median of $\log(M/M_{\odot})=7.5_{-0.2}^{+0.2}$, consistent with the range of stellar masses where $f_{\rm esc}$ is highest on average, based on the cosmological simulations of \citet{Ma2020}.
However, we note that selecting sources with high-\fesc{} based on extremely blue UV slopes preferentially identifies galaxies with the least impact from dust, which may bias our sample toward lower masses.
Thus, the typically low masses found for our sample does not necessarily indicate that somewhat higher mass galaxies cannot also have significant escape fractions.

\begin{figure}
    \centering
    \includegraphics[width=1.0\linewidth]{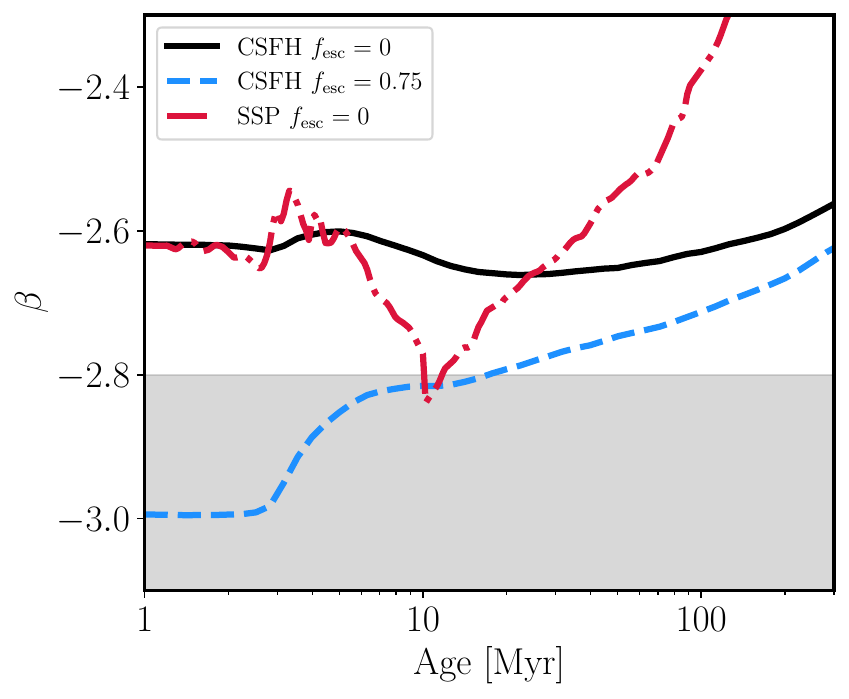}
    \caption{Evolution of the UV slopes as a function of galaxy age calculated from SED models constructed using BEAGLE. Each model is calculated assuming a $\log(U)=-2.5$, metallicity of $10\%Z_{\odot}$, and no dust attenuation. We present UV slopes for a CSFH at \fesc{}$=0$ (black line), \fesc{}$=0.75$ (blue dashed line), and an SSP with \fesc{}$=0$ (red dash-dotted line). We shade the region of of the figure ($\beta<-2.8$) coincident with our extremely blue galaxy sample.}
    \label{fig:burstbeta}
\end{figure}

\begin{figure*}
    \centering
    \includegraphics[width=1.0\linewidth]{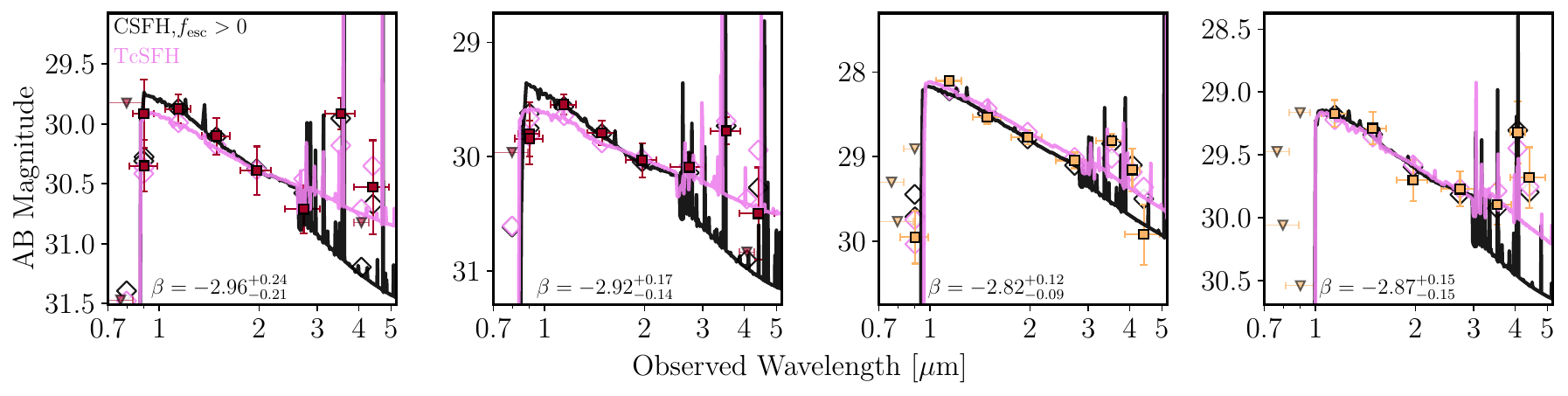}
    \caption{Comparison of best-fit SEDs when modelled as a density-bounded HII region with a CSFH (black) and a two-component SFH characterized by a recent cessation of star formation (pink). The photometric fluxes predicted from each of the models are shown as the empty diamonds. The observed fluxes and uncertainties are displayed as the red and orange points for objects that are F775W and F090W dropouts, respectively. }
    \label{fig:sfhmodel}
\end{figure*}

We explore the sizes of our extremely blue sample to investigate the geometrical and morphological characteristics of this candidate 
population of leakers.
Sizes were derived from the \jwst{}/NIRCam mosaics for each galaxy in the filter covering rest-frame 1500\angstrom{} using the method outlined in \citet{Chen2022}.
We utilize the inferred sizes to quantify the SFR surface densities ($\Sigma_{\rm SFR}=\rm SFR/2\pi r_e^2$), which are often thought to yield favorable conditions for \fesc{} at high $\Sigma_{\rm SFR}$ \citep[e.g.,][]{Sharma2017}.
The measured sizes and SFRs derived from the SED modelling for our extremely blue sample are shown in Figure~\ref{fig:sfrdensity}.
A majority (25/44) of the sample is unresolved implying compact sizes of $\leq160$ pc, while the remaining objects (19/44) have a median effective radius of $326$ pc, and up to $800$ pc for the most extended object.
Among the resolved galaxies, we find a median $\Sigma_{\rm SFR}$ of $1.4\rm~M_{\odot}/yr/kpc^2$. The remaining unresolved galaxies are all limited to above this value, in agreement with results at $z>10$ which have found values of $15-180\rm~M_{\odot}/yr/kpc^2$ \citep{Robertson2023}.
We compare our results to galaxies with confirmed LyC leakage at lower redshift including the Low-Redshift Lyman Continuum Survey \citep[LzLCS;][]{Flury2022a,Flury2022b} and Keck Lyman Continuum Survey \citet[KLCS;][]{Pahl2021}.
These results illustrate that while many objects with $\Sigma_{\rm SFR}>10\rm ~M_{\odot}/yr/kpc^2$ have $f_{\rm esc}>0$, this does not appear to be a prerequisite for leakage.
Our extremely blue sample displays a similar dynamic range of SFR surface densities, although they are systematically offset toward lower SFRs and more compact sizes.
Data that are able to probe smaller physical scales (such as through gravitational lensing) will be required to measure the SFR surface densities of this class of very blue galaxies.
Even among the resolved objects, the extremely blue UV slopes may be localized to a leaking region within the galaxy, such as that of the Sunburst Arc \citep{Kim2023}. In these cases, higher resolution from lensing is still required to better link the measured very blue UV slope and associated SFR surface density.

\subsection{The influence of bursty SFHs on very blue UV slopes}
\label{sec:tcsfh}

It is commonly assumed that Lyman continuum leakage is the primary physical  
interpretation driving extremely blue ($\beta<-2.8$) UV slopes in star forming galaxies. Here we consider whether 
short-term variations in the instantaneous star-formation rate may provide an alternative explanation. Theoretical studies have indicated that `bursty' star-formation histories are likely common in galaxies at early times \citep[e.g.,][]{Faucher2018, Tacchella2020, Ma2020, Furlanetto2022, Mirocha2023, Dome2023}. Observations with {\it JWST} are providing new evidence that this may be the case
\citep{Endsley2023,Dressler2023,Looser2023,Strait2023,Tacchella2023}. Shortly following the cessation of star formation after a burst, the overall ionizing photon production falls.  As a result, the nebular continuum strength decreases, allowing the stellar continuum to dominate in the UV. This enables 
very blue UV slopes to be observed.  In what follows, we explore the possibility of this physical picture as an avenue to describe the observed SEDs of our extremely blue sample, and discuss the impact on our interpretation of these sources.

We first explore whether UV slopes can reach sufficiently blue values after a burst of star formation. 
We quantify the time evolution of UV slopes using mock SEDs constructed with BEAGLE.
In Figure~\ref{fig:burstbeta} we display the expected UV slopes for simple stellar population (SSP) models, compared to models assuming a CSFH calculated at \fesc{}$=0$ and $0.75$.
After a time lag of $\sim5$ Myr following the cessation of star formation, the reduction of nebular continuum becomes significant, resulting in bluer UV slopes compared to the CSFH models.
The UV slopes continue to decrease up to an age of 10 Myr, reaching a minimum $\beta$ of $-2.8$. 
Following this minimum, the population becomes increasingly composed of less massive stars which have intrinsically redder UV spectra.
Thus, for ages of $\simeq5-30$ Myr, the burst models are able to produce UV slopes bluer than those possible with the CSFH models alone.
Based on these results, we find that the UV slopes of an SSP can reach the very blue value similar to that of a density-bounded model  with an \fesc{}$=0.75$. However this only occurs for a short period of time after the burst.

We apply this physical picture to our sample of very blue galaxies using models with a two-component star-formation history (TcSFH) such as those described in \citet{Endsley2023}.
Briefly, these models decouple the SFR over the most recent $20$ Myr from previous star formation that may be present, while keeping the remaining model parameters the same as our model setup described in Section~\ref{sec:properties}.  This allows for both upturns and lulls in recent star formation. Systematically, modelling the extremely blue sources with the TcSFH models does result in better fits to the observed data when compared to the CSFH models with \fesc{}$=0$.

Figure~\ref{fig:sfhmodel} compares the best-fit SEDs derived assuming a CSFH with non-zero escape fraction to the TcSFH model.
Among these examples, it is clear that the UV continua of galaxies with $\beta\simeq -2.8$ can be fit well by the TcSFH models, however the objects with bluer UV slopes are still challenging to reproduce without invoking LyC photon escape.  It is clear from the SEDs shown in Figure~\ref{fig:sfhmodel} that  the TcSFH models indicate significantly stronger rest-optical continuum levels than the density-bounded models. 
This in turn leads to 
the TcSFH stellar masses being systematically larger than those derived from the density-bounded models.
Quantitatively, we find an increase in stellar mass of $0.3$ dex on average, though the most extreme cases lead to an increase of $0.8$ dex. 
Ultimately, deep spectroscopy capable of detecting the continuum level in these sources should help distinguish between these scenarios for the very blue UV slopes.

%
%
\section{Summary}
\label{sec:summary}
In this paper we have leveraged deep \jwst{}/NIRCam imaging from the JADES program to explore the UV slopes for a statistical sample of galaxies up to $z\sim14$.
This sample covers a wide range of redshifts ($z\sim5-14$) and UV luminosities (\muv$=-22$ to $-16$), allowing us to explore trends between UV slopes and galaxy properties at high redshift.
Furthermore, we leverage both the wide areas and deep exposure times of JADES to identify rare $\beta\simeq-3$ galaxies, providing insight into some of the most extreme objects in the reionization era.
We summarize our main results below.

(i) We measured UV slopes for $1276$ galaxies identified within deep \jwst{}/NIRCam imaging from JADES, comprising four dropout samples spanning a range of redshifts from $z\simeq5$ to $z\simeq 14$.
The UV slopes of these high-redshift galaxies are typically blue, with median values of $\beta=-2.26^{+0.03}_{-0.03}$, $-2.32^{+0.03}_{-0.02}$, $-2.35^{+0.04}_{-0.04}$, and $-2.48^{+0.12}_{-0.14}$ at a redshift of $z\sim5.9$, $z\sim7.3$, $z\sim9.4$, and $z\sim12.02$, respectively.
We divide each of the three lower-redshift dropout samples into bins of \muv{} and explore how the typical UV slopes vary with UV luminosity.
This test reveals that throughout the redshift range sampled here, galaxies with lower UV luminosities typically display bluer UV slopes.
Using a set of simple analytic simulations we find that this trend is not explained by biases resulting from reduced efficiency of selecting redder galaxies.

(ii) At fixed UV luminosity we find that the average UV slopes become bluer in higher-redshift samples.
We find that the rate of this evolution is slowest (fastest) at low (high) UV luminosities.
Galaxies at $z>9$ and \muv{}$=-18$ display only a minor UV slope differences compared to the $z\simeq5-9$ population, while at an \muv{}$=-20$ we find significant evolution toward bluer UV slopes corresponding to $\Delta \beta=-0.15$.
The asymptotic behavior of typical UV slopes at lower luminosities suggests that the population is becoming increasingly saturated by galaxies approaching the intrinsic UV slope limit.
At yet higher redshift, we find increased evolution toward bluer UV slopes. 
We find that galaxies at $z\sim12$ display a typical UV slope of $\beta=-2.48^{+0.14}_{-0.12}$, compared to the $z\simeq9$ population which are just $\beta=-2.33^{+0.09}_{-0.08}$.
This implies that galaxies at the highest redshifts are subject to very little impact from dust.

(iii) The total fraction of galaxies with moderately red UV slopes ($\beta>-2$) decreases with redshift, and sharply falls of at $z\gtrsim 11$.
Such objects comprise $22.8\%$, $17.5\%$, and $19.4\%$ of the population at $z\simeq 5.86$, $z\simeq 7.28$, and $z\simeq 9.41$, respectively.
In contrast, the contribution of such sources to the highest redshift sample at $z\sim12$ is only $5\%$, as only one object falls into this category.
Notably, a similar decrease at $z\gtrsim12$ is found among galaxies selected based primarily on their photometric redshifts, suggesting that selection effects are not the principal cause for this evolution.

(iv) We identify a sample of 44 galaxies with extremely blue UV slopes ($\beta<-2.8$) that we classify as robust based on a set of criteria developed to maximize the likelihood that they are genuine.
This sample is composed of galaxies spanning redshifts of $z_{\rm phot}=5.42-10.96$ and UV luminosities corresponding to \muv{}$=-19.7$ to $-17.2$, and comprise $3.4\%$ of the total sample.
We find that the SEDs of this extremely blue sample are well-described by models of density-bounded HII regions with an average \fesc{} of $0.51$.

(v)  We search for evidence that the extremely blue sample display weak emission lines, which is expected if they exhibit high fractions of ionizing photon leakage.
We use color excesses to infer emission line strengths of our sample, and find that the median F356W-F444W color of our extremely blue sample is $0.19$ mag, which is systematically below the median of $0.39$ mag found for the full sample.
This implies that the extremely blue objects have weaker lines on average than the typical galaxies at this redshift, which is in line with the high values of \fesc{} inferred for the sample.
Model fits to this sample imply that they are characterized with very high sSFRs (median $79~ \rm Gyr^{-1}$), which are systematically elevated relative to the full sample of galaxies selected here (median $44~\rm Gyr^{-1}$).
Furthermore, we infer stellar masses of the extremely blue sample in the range $\log(\rm M/M_{\odot})=7.0-8.5$ (median $\log(\rm M/M_{\odot})=7.5\pm0.2$), which is consistent with simulation predictions for leakers.

(vi) Finally, we explore the SEDs of the extremely blue sample in context of models where recent star formation is decoupled from past star formation.
We find that the cessation of star formation is able to produce UV slopes bluer than that of CSFH models with \fesc{}$=0$ on short time scales, and provide a better fit to the data on average.
However the $\chi^2$ of the complex SFH models are not as good as those of the high \fesc{} models.
The resulting model fits imply stellar masses in excess of the high \fesc{} models by $0.3$ dex on average, with a maximum offset of $0.8$.
However in many cases, the complex SFHs require significant past star formation which have a stronger rest-optical continuum level relative to the density-bounded models and observed photometry.

Using JWST, we have constructed statistical samples of high-redshift galaxies, unveiling the properties of typical galaxies at $z\gtrsim9$, in addition to identifying rare and extreme galaxies.
Future campaigns providing spectroscopy of these sources will be a crucial step forward unveiling the nature of these galaxies with extremely blue UV slopes ($\beta<-2.8$).
Such data will confirm the UV slopes of the extremely blue systems, and will constrain the emission line EWs in a way that does not rely on model assumptions.
Further analysis of their spectra can shed light on the physical processes driving their extremely blue colors, and
will cement their place among galaxies responsible for cosmic reionization.

\section*{Acknowledgements}
MWT acknowledges support from the NASA ADAP program through the grant number 80NSSC23K0467.
DPS acknowledges support from the National Science Foundation through the grant AST-2109066.
LW acknowledges support from the National Science Foundation Graduate Research Fellowship under Grant No. DGE-2137419.
BDJ, BER, EE, DJE, \& CNAW  acknowledge JWST/NIRCam contract to the University of Arizona, NAS5-02015
WB, RM, \& JW acknowledge support by the Science and Technology Facilities Council (STFC), ERC Advanced Grant 695671 "QUENCH".
AJB \& JC acknowledge funding from the "FirstGalaxies" Advanced Grant from the European Research Council (ERC) under the European Union’s Horizon 2020 research and innovation programme (Grant agreement No. 789056)
S.C acknowledges support by European Union’s HE ERC Starting Grant No. 101040227 - WINGS.
ECL acknowledges support of an STFC Webb Fellowship (ST/W001438/1)
DJE is supported as a Simons Investigator.
RM acknowledges support by the UKRI Frontier Research grant RISEandFALL. RM also acknowledges funding from a research professorship from the Royal Society.
The research of CCW is supported by NOIRLab, which is managed by the Association of Universities for Research in Astronomy (AURA) under a cooperative agreement with the National Science Foundation.
JW acknowledges support from the Fondation MERAC.
The authors acknowledge use of the lux supercomputer at UC Santa Cruz, funded by NSF MRI grant AST 1828315. This material is based in part upon High Performance Computing (HPC) resources supported by the University of Arizona TRIF, UITS, and Research, Innovation, and Impact (RII) and maintained by the UArizona Research Technologies department.

\section*{Data Availability}
A portion of the observations utilized in this work can be accessed online at \href{https://archive.stsci.edu/hlsp/jades}{DOI: 10.17909/8tdj-8n28}.
Remaining data underlying this article will be shared on reasonable request to the corresponding author.

\bibliographystyle{mnras}
\bibliography{main}

\end{document}